\documentclass[prd,aps,floats,floatfix,eqsecnum,nofootinbib]{revtex4}
\usepackage{amsmath,amssymb,verbatim,epsfig,psfig,graphicx,rotating}
\usepackage{psfrag}
\newcommand{\be}{\begin{equation}}
\newcommand{\ee}{\end{equation}}
\newcommand{\bea}{\begin{eqnarray}}
\newcommand{\eea}{\end{eqnarray}}
\begin{document}
\title{Clarifying Inflation Models: the Precise Inflationary Potential from 
Effective Field Theory and the WMAP data}
\author{\bf D. Cirigliano$^{(a)}$}
\author{\bf H. J. de Vega$^{(b,a)}$}\email{devega@lpthe.jussieu.fr}
\author{\bf N. G. Sanchez $^{(a)}$}\email{Norma.Sanchez@obspm.fr}
\affiliation{$^{(a)}$
Observatoire de Paris, LERMA, Laboratoire Associ\'e au CNRS UMR 8112,
 \\61, Avenue de l'Observatoire, 75014 Paris, France. %}
\\$^{(b)}$ LPTHE, Laboratoire Associ\'e au CNRS UMR 7589,\\
Universit\'e Pierre et Marie Curie (Paris VI) et Denis Diderot (Paris VII),\\
Tour 24, 5 \`eme. \'etage, 4, Place Jussieu, 75252 Paris, Cedex 05,
France.}
\begin{abstract}
We clarify inflaton models by considering them 
as effective field theories in the Ginzburg-Landau spirit.
In this new approach, the precise form of the inflationary potential 
is constructed from the present WMAP data, and a useful scheme is prepared to 
confront with the forthcoming data. In this approach, 
the WMAP statement excluding
the pure $ \phi^4 $ potential implies the presence of an inflaton mass term
at the scale $ m \sim 10^{13}$GeV.
Chaotic, new and hybrid inflation models are studied in an unified way.
In all cases the inflaton potential takes the form
$ V(\phi) =  m^2 \;  M_{Pl}^2 \; v(\frac{\phi}{M_{Pl}}) $, where
all coefficients in the polynomial $ v(\varphi) $ are of order one.
If such potential corresponds to supersymmetry breaking, the
corresponding susy breaking scale is  $ \sqrt{m \;M_{Pl}} \sim 10^{16}$GeV
which turns to coincide with the grand unification (GUT) scale. 
The inflaton mass is therefore given by a see-saw formula 
$ m \sim M_{GUT}^2/M_{Pl} $.
The observables turn to be two-valued functions: one branch corresponds
to new inflation and the other to chaotic inflation, the branch point
being the pure quadratic potential. For red tilted spectrum, the potential
which fits the best the present data ($ |1-n_s| \lesssim 0.1 , 
\; r \lesssim 0.1 $) and which best prepares the way for
the forthcoming data is a trinomial polynomial with negative quadratic term
(new inflation). For blue tilted spectrum, hybrid inflation turns to be 
the best choice. In both cases we find an analytic formula relating the 
inflaton mass with the ratio $ r $ of tensor to scalar 
perturbations and the spectral index $ n_s $ of scalar perturbations: 
$ 10^6 \; \frac{m}{M_{Pl}} = 127 \; \sqrt{r|1-n_s|} $ where the numerical 
coefficient is fixed by the WMAP amplitude of adiabatic perturbations. 
Implications for string theory are discussed. 
\end{abstract}
\date{\today}
\maketitle
\tableofcontents

\section{Introduction}
Inflation was originally proposed to solve several outstanding
problems of the standard Big Bang model
\cite{guth,kolb,coles,lily,riottorev} thus becoming an
important paradigm in cosmology. At the same time, inflation
provides a natural mechanism for
the generation of scalar density fluctuations that seed large
scale structure, thus explaining the origin of the temperature
anisotropies in the cosmic microwave background (CMB)\cite{cobe},  as well as
the tensor perturbations (primordial gravitational waves). Recently,
the Wilkinson Microwave Anisotropy Probe (WMAP) collaboration has
provided a full-sky map of the temperature fluctuations of the
cosmic microwave background (CMB) with unprecedented accuracy and
an exhaustive analysis of the data confirming the basic and robust
predictions of inflation\cite{WMAP}.

During inflation quantum vacuum fluctuations are generated with
physical wavelengths that grow faster than the Hubble radius, when
the wavelengths of these perturbations cross the horizon they
freeze out and decouple\cite{kolb,lily,riottorev}.
Wavelengths that are of cosmological relevance today  re-enter the
horizon during the matter dominated era when the scalar (curvature)
perturbations induce temperature anisotropies imprinted
on the CMB at the last scattering surface\cite{pert,hu}. Generic
inflationary models predict that these are mainly Gaussian and
adiabatic perturbations with an almost scale invariant spectrum.
These generic predictions are in spectacular agreement with the CMB
observations as well as with a variety of large scale structure data
\cite{WMAP}. The WMAP data \cite{WMAP} clearly display an anti-correlation
peak in the temperature-polarization (TE) angular power spectra at
$l\sim 150$, providing one of the most striking confirmations of adiabatic
fluctuations as predicted by inflation\cite{WMAP}. 

The classical dynamics of the inflaton (a massive scalar field) 
coupled to a cosmological background clearly shows that inflationary
behaviour is an {\bf attractor} \cite{bgzk}. This is a generic and robust
feature of inflation. 
The robust predictions of inflation (value of the  entropy of the universe, 
solution of the flatness problem, small adiabatic Gaussian density 
fluctuations explaining the CMB anisotropies, ...) which are common to many 
available inflationary scenarios, show the predictive power of the 
inflationary paradigm. Whatever the microscopic model for the early universe 
(GUT theory) would be, it should include inflation with the generic features 
we know today. 

Inflationary dynamics is typically studied by treating  the
inflaton as a homogeneous  classical scalar
field\cite{kolb,coles,lily} whose evolution is determined
by a classical equation of motion, while the inflaton 
quantum fluctuations (around the classical value and in the
Gaussian approximation) provide the seeds for the scalar
density perturbations of the metric. In quantum field theory,
this classical inflaton corresponds to the expectation value 
of a quantum field operator in a translational invariant state.
Important aspects of the inflationary dynamics, as resonant
particle production and the nonlinear back-reaction that it generates,
require a full quantum treatment of the inflaton for their consistent
description. The quantum dynamics of the inflaton 
in a non-perturbative framework and its consequences on the CMB anisotropy 
spectrum were treated in refs.\cite{cosmo,cosmo2}. Particle decay in de Sitter 
background and during slow roll inflation is studied in ref.\cite{desin} 
together with its implication for the decay of the density fluctuations. 

Inflation as known today should be considered as an {\bf effective theory},
that is, it is not a fundamental theory but a theory of a
condensate (the inflaton field) which follows from a more fundamental one 
(the GUT model). 
The inflaton field $ \phi $ may {\bf not}
correspond to any real particle (even unstable) but is just an {\bf effective}
description while the microscopic description should come from the GUT model. 
At present, there is no derivation of the inflaton model from
the microscopic GUT theory. However, the relation of inflation to the GUT
theory is like the relation of the effective Ginzburg-Landau theory of 
superconductivity with the microscopic BCS theory. Or like the relation of the 
$O(4)$ sigma model, an effective particle theory for low energy, with the 
microscopic quantum chromodynamics (QCD). 

The aim of this paper is to provide a clear understanding of inflation
and the inflaton potential from effective field theory and the 
WMAP data. This clearly places inflation within the perspective and
understanding of
effective theories in particle physics. In addition, it sets up a clean way to 
directly confront the inflationary predictions with the forthcoming CMB data 
and select a definitive model.

The following inflaton potential or alternatively the hybrid inflation model 
are rich enough to describe the physics
of inflation and accurately reproduce the available data \cite{WMAP}:
\be\label{potint}
V(\phi) =  |m^2| \;  M_{Pl}^2 \left[ v_0 \pm \frac12 \; \varphi^2 + 
\frac23 \; \gamma \; 
\varphi^3 + \frac1{32} \; \kappa  \; \varphi^4 \right] \; .
\ee
Here $ \varphi \equiv \frac{\phi}{M_{Pl}} \; , |m| \sim 10^{13}$GeV, 
the dimensionless parameters 
$ \gamma $ and $ \kappa $ are of order one, and $ v_0 $ is  such that
$ V(\phi) $ and $ V'(\phi) $ vanish at the absolute minimum of $ V(\phi) $.
This ensures that inflation ends after a finite time
with a finite number of efolds. $ \kappa $ must be positive to ensure stability
while $ \gamma $ and the mass term $ \varphi^2 $ can have either sign. 
$ \gamma $ describes how asymmetric is the potential while  $ \kappa $
determines how steep it is.
Notice that there is {\bf no fine tuning} here once the mass scale $ |m| $ 
is fixed.

The potential eq.(\ref{potint}) cover a wide class of inflationary scenarios:
small field scenarios (new inflation) for spontaneously broken symmetric
potentials (negative mass square), as well as large field scenarios
(chaotic inflation) for unbroken symmetric potentials (positive mass square).
Coupling the inflaton to another scalar field yields the hybrid type scenarios.

In the context of an effective theory or Ginzburg-Landau model it is {\bf 
highly unnatural} to drop the quadratic term $ \varphi^2 $. 
This is to  exactly choose the  {\bf critical} point of the model
$ m^2 = 0 $. In fact, the recent WMAP \cite{WMAP} 
statement unfavouring the monomial  $ \varphi^4 $ potential just supports a 
generic polynomial inflaton potential as in eq.(\ref{potint}).
Excluding the quadratic mass term in the potential 
$ V(\phi) $ implies to fine tune to zero the mass term which is 
only justified at isolated (critical) points. 
Therefore, from a physical  point of view, the pure quartic potential 
$ \varphi^4 $ is a weird choice implying to fine tune to zero the coefficient of 
$ \varphi^2 $.

We obtain analytic and unifying expressions for chaotic and new inflation
for the relevant observables:
the amplitude for scalar fluctuations $ |{\delta}_{k\;ad}^{(S)}|^2 $, 
spectral index  $ n_s $ and ratio $ r $ of tensor to scalar
perturbations as well as for hybrid inflation and plot them for the three
scenarios. Particularly interesting are the plots of $ n_s $ vs. $ r $ 
(figs. 7, 18, 19, 20 and 21). 

\bigskip

We express the ratio of the inflaton mass and the Planck mass 
$ x \equiv 10^6 \; \frac{m}{M_{Pl}} $ in terms
of the amplitude of adiabatic perturbations and the parameters in the 
potential. Furthermore,  we can express $ x $
in terms of observable quantities as $ r $ and $ n_s $. 
We find for new inflation when both $ r $ and $ | n_s - 1 | $ are small,
\be \label{xint}
x = 5 \, \pi \, \sqrt{3} \; 10^5 \, |{\delta}_{k\;ad}^{(S)}| \; 
\sqrt{r(1-n_s)} = 127 \; \sqrt{r(1-n_s)} \pm 6 \% \; .
\ee
where the $ \pm 6 \% $ correspond to the error bars in the 
amplitude of adiabatic perturbations\cite{WMAP}.
From figs. \ref{trimasamulti}, \ref{trinsmulti} and \ref{trirmulti}
we can understand how the mass ratio $ \frac{m}{M_{Pl}} $ varies with $ n_s $ 
and $ r $. We find a {\bf limiting} value  $ x_0 \equiv 10^6  \; 
\frac{m_0}{M_{Pl}} \simeq 0.1 $ for the inflaton mass 
such that $ m_0 \simeq  10^{-7} \; M_{Pl} $ is a {\bf minimal} inflaton mass 
in order to keep $ n_s $ and $ r $ within the WMAP data.

\bigskip

New inflation arises for broken symmetric potentials (the minus sign in front
of the $ \varphi^2 $ term) while chaotic inflation appears both for unbroken
and broken symmetric potentials. 
For broken symmetry, we find that analytic continuation
connects the observables for chaotic and new inflation:
the observables 
are {\bf two-valued} functions of $ y \equiv \kappa \, N $. 
($N$ being the number of efolds from the first horizon 
crossing to the end of inflation).
One branch corresponds to new inflation and the other 
branch to chaotic inflation.
As shown in figs. 4-7, 9, 12 and 15, $n_s, \; r $ and 
$ |{\delta}_{k\;ad}^{(S)}|^2 $
for chaotic inflation are connected by analytic continuation to
the same quantities for new inflation. The branch point
where the two scenarios connect 
corresponds to the monomial $ +\frac12 \, 
\varphi^2 $ potential ($\kappa=\gamma=0$). 

\bigskip

The potential which {\bf best fits} the present data for a red tilted 
spectrum ($ n_s < 1 $)
and which {\bf best prepares} the way to the expected data (a small 
$ r \lesssim 0.1 $) is given by the trinomial potential eq.(\ref{potint}) with
a negative  $ \varphi^2 $ term, that is {\bf new} inflation. 

In new inflation we have the upper bound 
$$ 
r \leq \frac{8}{N} \simeq  0.16 \quad  .
$$
This upper bound is attained by the quadratic monomial. 
On the contrary, in chaotic inflation for both signs of the $ \varphi^2 $ term,
$ r $ is bounded as
$$
0.16 \simeq \frac8{N} < r < \frac{16}{N}  \simeq 0.32\quad ,
$$
This bound holds for all values of the cubic coupling $ \gamma $ which describes
the asymmetry of the potential. 
The lower and upper bounds for $ r $ are saturated by the quadratic and quartic
monomials, respectively.

\bigskip 

For chaotic and new inflation, we find the following properties:
\begin{itemize}
\item{ $n_s$ is bounded as
$$ 
n_s \leq 1 - \frac2{N} \simeq 0.96 \qquad  \mbox{chaotic~inflation,}\qquad 
n_s \leq 1 - \frac{1.558005\ldots}{N} \simeq 0.9688 \qquad  
\mbox{new~inflation} \; .
$$ 
The value at the bound for chaotic inflation corresponds to the  quadratic monomial.}
\item{ $n_s$ decreases with the steepness $ \kappa $ 
for fixed asymmetry $ h \equiv\gamma \; \sqrt{\frac{8}{\kappa}} < 0 $ and grows with 
the asymmetry $ |h| $ for fixed steepness $ \kappa $.}
\end{itemize}

For chaotic inflation $ r $  grows with the steepness $ \kappa $ for fixed
asymmetry $ h < 0 $ and decreases with the asymmetry $ |h| $ for fixed
steepness $ \kappa $. Also, in chaotic inflation $ r $ decreases with $ n_s $.

For new inflation  $ r $ does the opposite: {\bf it decreases} with
the steepness $ \kappa $ for fixed asymmetry $ h < 0 $ while it grows 
with the asymmetry $ |h| $ for fixed  steepness 
$ \kappa $. Also, in new inflation $ r $ grows with $ n_s $.

All this is valid
for the general trinomial potential eq.(\ref{potint}) and can be
seen in figs. 7, 18, 20 and 21.  In addition, $ r $ {\bf decreases} for 
increasing  asymmetry $ |h| $ at a fixed $ n_s $ in new inflation 
(with $ h<0 $). As a consequence, the trinomial 
potential  eq. (\ref{potint}) can yield {\bf very small $r$} for red tilt with
{\bf $ n_s < 1 $ and near unit} for new inflation.  

\bigskip

Hybrid inflation always gives a blue tilted spectrum $ n_s > 1 $ in the
$\Lambda$-dominated regime, 
allowing $ n_s - 1 $ and $ r $ to be small. Interestingly enough, 
we obtain for hybrid inflation a formula for the mass ratio $ x $
with a similar structure to eq.(\ref{xint}) for new inflation:
$$
x =  10^6 \; \frac{m}{M_{Pl}} = 127 \, 
\sqrt{r \, \left(n_s -1 + \frac38 \; r\right) } \quad .
$$
This is plotted in fig. 22-23 showing that $ \frac{m}{M_{Pl}} $ 
{\bf decreases} when $ r $  {\bf and} $ n_s - 1 $ both approach zero.
We relate the cosmological constant in the hybrid inflation Lagrangian
with the ratio $r$ as 
$$
\frac{\Lambda_0}{M^4_{Pl}} = 0.329 \times 10^{-7} \; r \; ,
$$
and we find that $ (n_s - 1) $ gives
an {\bf upper bound} on the cosmological constant:
$$
  \frac{\Lambda_0}{m^2 \;M_{Pl}^2} < \frac2{n_s-1} \; .
$$

In order to reproduce the CMB data, the inflationary potentials 
in the slow roll scenarios considered in this article must
have the structure
$$
V(\phi) =  M^4 \; v\!\left(\frac{\phi}{M_{Pl}}\right) \; ,
$$
where $ v(0) = v'(0) = 0 $  and all higher derivatives at the origin are of the
{\bf order one}. The inflaton mass is therefore given by a see-saw-like
formula
\be \label{mint}
m \simeq \frac{M^2}{M_{Pl}} \; .
\ee
As stated above, the WMAP data imply $ m \sim 10^{13}$GeV, Eq. (\ref{mint})
implies that $M$ is {\bf precisely} at the grand unification 
scale $M \sim 10^{16}$GeV \cite{kolb,coles,lily}. 
Three strong independent indications
of this scale are available nowadays: 1) the convergence of the 
running electromagnetic, weak and strong couplings, 2) the large mass scale
to explain the neutrino masses via the see-saw mechanism and 3) the
scale $M$ in the above inflaton potential. Also, notice that eq.(\ref{mint}) 
has the structure of the moduli potential coming from supersymmetry breaking.
Therefore, the supersymmetry breaking scale would be at the GUT scale too.

\bigskip

In order to generate inflation in string theory, one needs first to generate
a mass scale like $ m $ and $ M_{GUT} $ related by eq.(\ref{mint}). Without 
such mass scales there is {\bf no} hope to generate a realistic cosmology
reproducing the observed CMB fluctuations. However, an effective description
of inflation in string theory could be at reach \cite{noscu}

\bigskip

In summary, for small  $ r \lesssim 0.1 $ and $ n_s $ near unit,
{\bf new} inflation from the trinomial potential eq.(\ref{potint})
and {\bf hybrid} inflation emerge as the {\bf best} candidates.
Whether $n_s$ turns to be above or below unit will
choose hybrid or new inflation, respectively. 
In any case $|n_s -1|$ turns to be of order $ 1/N $ ($N$ being the number 
of efolds from the first horizon crossing to the end of inflation).
This can be understood intuitively as follows: the geometry of the universe
is scale invariant during de Sitter stage since the metric takes in conformal 
time the form 
$$
ds^2 = \frac1{(H \; \eta)^2}\left[ (d \eta)^2 - (d \vec x)^2 \right] \; .
$$
Therefore, the primordial power generated is scale invariant except
for the fact that inflation is not eternal and lasts for $N$  efolds.
Hence, the  primordial spectrum is scale invariant up to $ 1/N $ corrections.
Also, the ratio $ r $ turns to be of order  $ 1/N $ (chaotic and new inflation)
or  $ 1/N^2 $ (hybrid inflation).

\section{Inflation and the Inflaton Field: an Effective Field Theory}

Inflation is part of the standard cosmology since several years. 
Inflation emerged in the 80's as the only way to explain the `bigness'
of the universe, that is, the value of the entropy of the
universe today $ \sim 10^{90} \sim (e^{69})^3 $. Closely related to
this, inflation solves the horizon and flatness problem, thus explaining
the quasi-isotropy of the CMB.  For a recent outlook see \cite{dos}.

\bigskip

The inflationary era corresponds to the scale of energies of Grand
Unification. It is not yet known which field model appropriately
describes the matter at such scales. Fortunately, one does not need a
detailed description in order to investigate inflationary cosmology,
one needs the expectation value of the quantum energy density
($T_{00}$) which enters in the r. h. s. of the Einstein-Friedman
equation 
\be \label{ef}
\left[ \frac{1}{a(t)} \; \frac{da}{dt} \right]^2 = \frac{1}{3 \; M^2_{Pl}}
 \; \rho(t) \; ,
\ee
This is dominated by field condensates. Since
fermion fields have zero expectation values, only the bosonic fields
are relevant. Bosonic fields do not need to be fundamental
fields, they can describe fermion-antifermion  pairs $ < {\bar \Psi} \Psi> $
in a grand unified theory (GUT). In order to describe the cosmological 
evolution is enough to consider the effective dynamics of such condensates. 
In fact, one condensate field is enough to obtain inflation. 
This {\it condensate} field is usually called `inflaton' and its dynamics can 
be described by a Ginzburg-Landau Lagrangian in the cosmological background 
\begin{equation}\label{FRW} 
ds^2= dt^2-a^2(t) \; d{\vec x}^2 
\end{equation}
That is, an
effective local Lagrangian containing terms of dimension less or equal
than four in order to be renormalizable,
\be\label{lagra}
{\cal L} = a^3(t) \left[ \frac{{\dot \phi}^2}{2} - \frac{({\nabla
      \phi})^2}{2 \,  a^2(t)} - V(\phi) \right]
\ee
Here, the inflaton potential $ V(\phi) $ is often taken as a quartic
polynomial: $ V(\phi)= \frac{m^2}{2} \; \phi^2 + \frac{\lambda}{4}\;
\phi^4 $. 

\bigskip

The Einstein-Friedman equation (\ref{ef}) for homogeneous fields take the form 
\be\label{frih}
H^2(t) = \frac{1}{3 \; M^2_{Pl}}  \; \left[ \frac{{\dot \phi}^2}{2} 
+ V(\phi) \right] \quad , \quad H(t) \equiv \frac{1}{a(t)} \; \frac{da}{dt}
\ee
where we used eq.(\ref{lagra}), and $ H(t) $ stands for the Hubble parameter.

\medskip

The inflaton field $ \phi $ may {\it not} correspond to any real particle
(even unstable) but is just an {\bf effective} description of the
dynamics. The detailed microscopical description should be given
by the GUT. Somehow, the inflaton is to the  microscopic GUT theory
what the Ginzburg-Landau effective theory of superconductivity is to the
microscopic BCS theory. Another relevant example of 
{\it effective field theory} in particle physics is the $O(4)$ sigma model 
which describes quantum chromodynamics (QCD) at low energies \cite{quir}. 
The inflaton model can be thus considered as an {\bf effective theory}.
That is, it is not a fundamental theory but a theory on the condensate
(the inflaton field) which follows from a more fundamental theory 
(the GUT model), by integrating over the basic fields in the latter. 
In principle, it should be possible to derive the inflaton Lagrangian from a 
GUT model including GUT fermions and gauge fields. Although, such derivation 
is not yet available, one can write down, as in the case of the sigma model 
describing the low energy behaviour of QCD, the effective 
Lagrangian for the particles of interest (the pions, the sigma and photons)
without explicit calculation in the fundamental theory. The guiding 
principle being the symmetries to be respected by the effective model
\cite{quir}. 
Contrary to the sigma model where the chiral symmetry strongly constraints the 
model\cite{quir}, only Lorentz invariance can be imposed to the inflaton 
model. Besides that, one can always eliminate linear terms in the Lagrangian 
by a constant shift of the inflaton field.

Restricting ourselves to renormalizable theories we can choose a general 
quartic Lagrangian with
\be\label{V}
V(\phi)= V_0 + \frac{m^2}{2} \; \phi^2 + \frac{ |m| \; g }{3} \; \phi^3 + 
\frac{\lambda}{4}\; \phi^4 \; . 
\ee
where $ \lambda $ and $ g $ are dimensionless parameters and $V_0$ is chosen 
such that $V(\phi)$ vanishes at its absolute minimum. This ensures that 
inflation ends after a finite time with a finite number of efolds. We choose 
$ \lambda > 0 $ as a stability condition in order to have a potential bounded 
from below while $ m^2 $ and $ g $ may have any sign. An inflaton potential of 
this type was considered in ref.\cite{hbkp}.

As it is known, in order to reproduce the CMB anisotropies, one has to
choose $ m $ around the GUT scale $ m \sim 10^{-6} \; M_{Pl}  \sim 10^{13}$GeV,
and the coupling $ \lambda $ very small ($ \lambda \sim 10^{-12} $) 
\cite{kolb,coles,lily} while $g$ may be just omitted. 

Let us see that the choice $ \lambda \sim 10^{-12} $ is not independent from 
the value of $ m / M_{Pl} \sim 10^{-6} $. Let us define a dimensionless field 
$ \varphi \equiv \frac{\phi}{M_{Pl}} $, the potential $ V $ for $ m^2 > 0 $ 
takes now the form,
$$
V(\phi) = m^2 \;  M_{Pl}^2 \left[ \frac12 \; \varphi^2 + 
\frac13 \; g \;  \frac{M_{Pl}}{m} \; \varphi^3 +
\frac14 \; \lambda  \; \frac{M_{Pl}^2}{m^2} \; \varphi^4 \right] + V_0 \; , \; 
\varphi = \frac{\phi}{M_{Pl}}
$$
or
\be \label{vnat}
V(\varphi) = m^2 \;  M_{Pl}^2 \left[  v_0 + \frac12 \; \varphi^2 + 
\frac23 \; \gamma \; 
\varphi^3 + \frac1{32} \; \kappa  \; \varphi^4\right] \; .
\ee
Here,
\be \label{gaka}
\gamma \equiv  g \;  \frac{M_{Pl}}{2 \, m} \quad ,  \quad 
\kappa  \equiv 8 \, \lambda \; \frac{M_{Pl}^2}{m^2} \quad ,  \quad v_0  \equiv 
\frac{V_0}{m^2 \;  M_{Pl}^2} \; ,
\ee
are all three of  {\bf order one} in order to reproduce the CMB anisotropies. 
Hence, once the mass $m$ is chosen to be in the scale $ \sim 10^{13}$GeV, 
the remaining parameters $ \gamma , \; \kappa , \ldots $ turn out to be
of order one. In other words, {\bf there is no fine tuning} in the choice of 
the inflaton self-couplings. $ \gamma $ describes how asymmetric is the 
potential while  $ \kappa $ determines how steep it is.

In typical inflationary scenarios one  initially has $ V \sim H^2 \, M_{Pl}^2 $
 and $ H \sim 5 \, m $. This makes the parametrization $V(\varphi)$ as in 
eq.(\ref{vnat}) very natural with $ \varphi $ less than (or of the order) one 
at the beginning of inflation.

\bigskip

In the context of an effective theory or Ginzburg-Landau model it is highly 
unnatural to set $ m=0 $. This corresponds to be exactly at the critical point 
of the model where the mass vanishes, that is, the correlation length is 
 infinite in the statistical mechanical context.
In fact, the recent WMAP \cite{WMAP} statement unfavouring the $ m= 0 $ 
choice (purely $ \phi^4 $ potential) just supports a generic 
polynomial inflaton potential possessing
a $ \phi^2 $ mass term plus  $ \phi^4 $ (plus eventually other terms).

We want to stress that excluding the quadratic mass term in the potential 
$ V(\phi) $ implies to fine tune to zero the mass term which is 
only justified at 
isolated points (a critical point in statistical mechanics). 
Therefore, from a physical  point of view, the pure quartic potential
is a weird choice implying to fine tune to zero the coefficient of the 
mass term. In other words, one would be considering a field with 
self-interaction but lacking of the mass term.

\medskip

Choosing $ g = 0 $ implies that $ \varphi \to -\varphi $ is a symmetry of the 
inflaton potential.
We do not see reasons based on fundamental physics to choose a zero or a 
nonzero $ g $.
Only the phenomenology, that is the fit to CMB data, can decide for the moment 
on the value of  $ g $. 

A model with only one field is clearly unrealistic since the inflaton would 
then describe a stable and ultra-heavy (GUT scale) particle. It is
necessary to couple the inflaton with lighter particles, then, the inflaton
can decay into them. 
There are many available scenarios for inflation. Most of them add
other fields coupled to the inflaton. This variety of inflationary
scenarios may seem confusing since several of them are compatible with the
observational data\cite{WMAP}. Indeed, future observations should
constraint the models more tightly excluding some families of them. Anyway, the
variety of acceptable inflationary models shows the {\bf power} of the
inflationary paradigm. Whatever the correct microscopic model for
the early universe would be, it should include inflation with the generic 
features we know today.
In addition, many inflatons can be considered (multi-field inflation). 
Such family
of models introduce extra features as non-adiabatic (isocurvature)
density fluctuations, which in turn become strongly constrained by the WMAP 
data \cite{WMAP}.

The scenarios where the inflaton is treated classically are usually
characterized into small and large fields scenarios. In small fields
scenarios the initial classical amplitude of the inflaton is assumed
small compared with $ M_{Pl} $, while in large field scenarios
the inflaton is initially of the order $\sim  M_{Pl} $ \cite{lily}. The 
first type of scenarios is usually realized with spontaneously broken symmetric
potentials ($ m^2 < 0 $, 'new inflation', also called `small field 
inflation'), while for the second type scenarios 
one can just use unbroken potentials ($ m^2 > 0 $, `chaotic inflation' also 
called `large field inflation').

We will restrict in this paper to inflationary potentials of degree four 
as in eq.(\ref{vnat}).
This ensures that the corresponding quantum theory is renormalizable.
Notice that being the inflaton model an effective theory nothing forbids 
to consider inflationary potentials of arbitrary high order. 
In general, the potential $V(\phi)$ will have the form:
\be \label{vgen}
V(\phi) =  m^2 \;  M_{Pl}^2 \; v\!\left(\frac{\phi}{M_{Pl}}\right) \; ,
\ee
where $ v(0) = v'(0) = 0 $  and all higher derivatives at the origin are of the
{\bf order one}. The arbitrary function $ v( \varphi ) $ allows detailed fits.
However, already a quartic potential is rich enough to describe the full 
physics and to reproduce accurately the data. In such case we have from 
eqs.(\ref{vnat}) and (\ref{vgen}) for $ m^2 > 0 $,
$$
v(\varphi) = v_0 + \frac12 \; \varphi^2 + \frac23 \; \gamma \; 
\varphi^3 + \frac1{32} \; \kappa  \; \varphi^4 \; .
$$
In dimensionless variables the Einstein-Friedman equation (\ref{frih}) takes 
the form,
\be\label{efsd}
h^2(\tau) = \frac{1}{3} \; \left[ \frac{{\dot \varphi}^2}{2} + 
v(\varphi) \right] \; ,
\ee
where we introduced the dimensionless time variable $ \tau \equiv m \; t , \;
{\dot \varphi} \equiv \frac{d \varphi}{d\tau} $ and 
$ h(\tau) \equiv \frac{H(t)}{m} $.

\medskip

The evolution equation for the field $ \varphi(\tau) $ then reads
\be\label{fi4sd}
{\ddot \varphi} + 3 \, h \, {\dot \varphi} + v'( \varphi ) = 0 \; .
\ee
In the case of the quartic potential eqs.(\ref{efsd})-(\ref{fi4sd})
become,
\bea
&&h^2(\tau) = \frac{1}{6} \; \left[ {\dot \varphi}^2 + \varphi^2 + 
\frac{2}{3}  \; \gamma \; \varphi^3 + \frac1{16} \; \kappa  \; \varphi^4 + 
2 \, v_0 \right] \; , \cr \cr
&&{\ddot \varphi} + 3 \, h \, {\dot \varphi} + \varphi +2 \; \gamma \; 
\varphi^2 + \kappa  \; \varphi^3 = 0 \; .
\eea
We similarly treat the spontaneous symmetry breaking case $ m^2 < 0 $
by setting 
\be
V(\phi) = |m|^2 \;  M_{Pl}^2 \; v(\varphi) \quad \mbox{where} \quad
 v(\varphi) = -\frac12 \; \varphi^2 + \frac23 \; \gamma \; 
\varphi^3 + \frac1{32} \; \kappa  \; \varphi^4 + v_0 \; .
\ee
and
\be \label{capa}
\gamma \equiv  g \;  \frac{M_{Pl}}{2 \, |m|} \quad ,  \quad \kappa  \equiv 
8 \, \lambda \; \frac{M_{Pl}^2}{|m^2|} \quad ,  \quad v_0  
\equiv \frac{V_0}{|m^2| \;  M_{Pl}^2} \; ,
\ee
instead of eqs.(\ref{vnat})-(\ref{gaka}).
In the case $ m^2 < 0 $ the minimum of the potential is not at
the origin but the second derivative of $ v(\varphi) $ at its absolute
minimum is always positive. That is, the physical mass square
of the inflaton field is positive whatever the sign of $ m^2 $.

\bigskip

We restrict ourselves to potentials $ V(\phi) $ which are polynomials of degree
four in $ \phi $. Higher order polynomials describe non-renormalizable 
interactions, although being acceptable as effective field 
theories \cite{quir}.
As we show below, already quartic polynomials for $ V(\phi) $ are rich enough 
to describe the full physics and fit the present data. 
One could use higher order polynomials to refine the fits. 

\section{Density fluctuations: chaotic inflation and new inflation}

We present here the density fluctuations for a general inflationary potential 
as in eq.(\ref{vnat}). Scalar and tensor perturbations in inflation
have been the subject of intense activity\cite{lily,riottorev,hu,pert} in 
particular, their contrast with the WMAP data \cite{WMAP,varios}

\subsection{The binomial  inflaton potential}

Let us start with the case where the potential is just a binomial in 
$ \varphi^2 $. We have for the unbroken symmetry case $ m^2 > 0 $,
\be \label{binonr}
 v_{ub}(\varphi) = \frac12 \; \varphi^2 + \frac1{32} 
\; \kappa  \; \varphi^4 \; .
\ee
and for the symmetry breaking case $ m^2 < 0 $,
\be \label{binor}
 v_{b}(\varphi) =\frac{\kappa}{32} \left(\varphi^2 - \frac{8}{\kappa} \right)^2
= -\frac12 \; \varphi^2 +\frac1{32} \; \kappa  \; \varphi^4 + 
\frac{2}{\kappa}\;.
\ee
where $ \kappa $ is defined by eq.(\ref{capa}).
The value of $v_0$ is chosen such that  $v=0$ 
at its absolute minimum ($ v_0 = 0 $ for $v_{ub}$ and 
$ v_0 = \frac{2}{\kappa} $
for $v_b$). This ensures that inflation ends after a finite 
time with a  finite number of efolds. 

\medskip

\begin{figure}[htbp]
{\epsfig{file=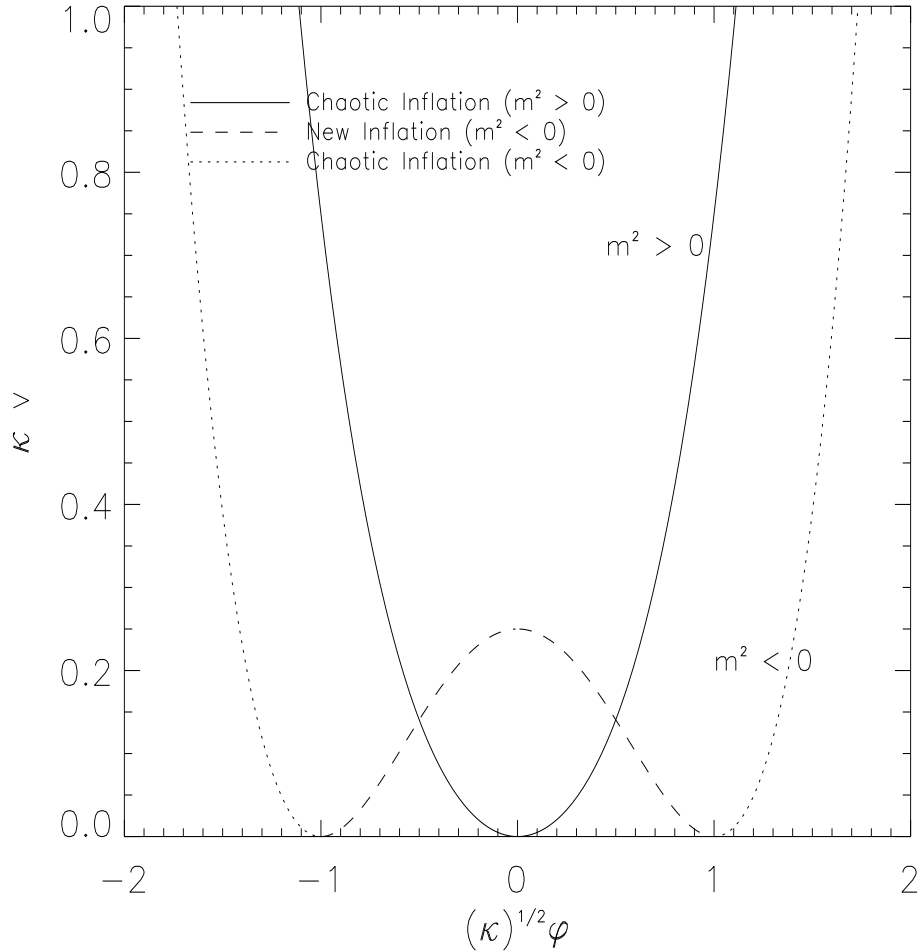,width=14cm,height=14cm}}
\caption{ The binomial potential eqs. (\ref{binonr})-(\ref{binor})
for both unbroken and broken cases: $ \kappa \;  v_{ub}(\varphi) 
\; ( m^2 > 0 )$ and
$ \kappa \;  v_b(\varphi) \; ( m^2 < 0 ) $ vs. $ \sqrt{\frac{\kappa}{8}} \; \varphi $.
} \label{potbino}
\end{figure}

The chaotic scenario is realized for $ m^2 > 0 $ with the inflaton starting
at some value $ \varphi $ of the order one, ($0 < \varphi < +\infty$). 
By the end of inflation $ \varphi $
is near the minimum of the potential at $ \varphi = \varphi_0 = 0 $
(see fig. \ref{potbino} ). 

\medskip

The new inflationary scenario is realized for  $ m^2 < 0 $ with the symmetry 
breaking potential  eq.(\ref{binor}) and the initial condition  $ \varphi $ 
very close to the origin $ \varphi = 0 , \; (0 < \varphi < \varphi_0 = 
\sqrt{\frac{8}{\kappa}})$,
where $\varphi_0 $  is  the minimum of the potential. 
By the end of  inflation, $ \varphi $ is near $ \varphi_0 $ (see fig. 
\ref{potbino} ).

In addition, one obtains chaotic inflation in the case $ m^2 < 0 $ choosing
the initial  $ \varphi $ larger than $ \varphi_0 , \; (\varphi_0<  
\varphi <  +\infty)$. 

We display in fig. \ref{potbino} the binomial potentials $ \kappa \;  
v_{ub}(\varphi) $ and $ \kappa \;  v_b(\varphi) $ as functions of 
$ \sqrt{\frac{\kappa}{8}} \; \varphi $.

\begin{figure}[htbp]
{\epsfig{file=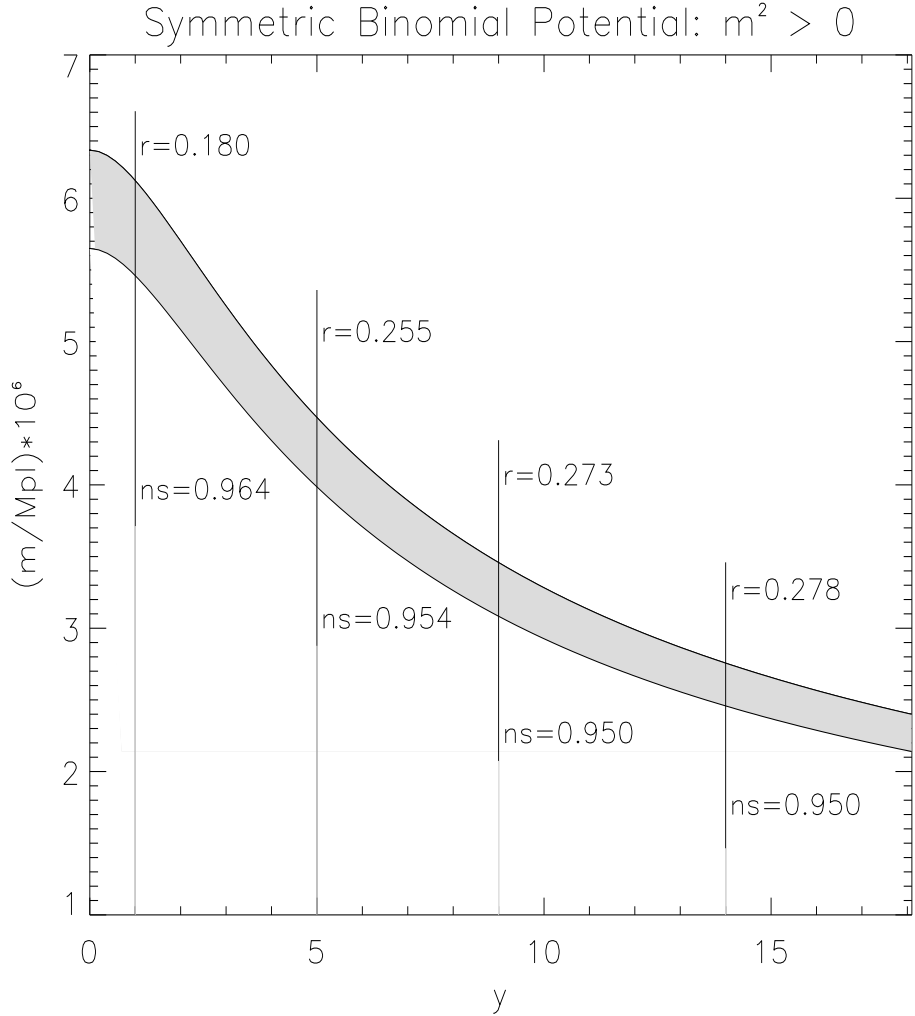,width=14cm,height=14cm}}
\caption{Upper and lower bounds for the mass ratio 
$10^6 \; m/M_{Pl}$ as functions of 
$ y = \kappa \, N $ for $m^2 > 0$ (chaotic inflation) for $ N = 60 $
with the binomial potential eq.(\ref{binonr}). } \label{xm2p}
\end{figure}

\begin{figure}[htbp]
{\epsfig{file=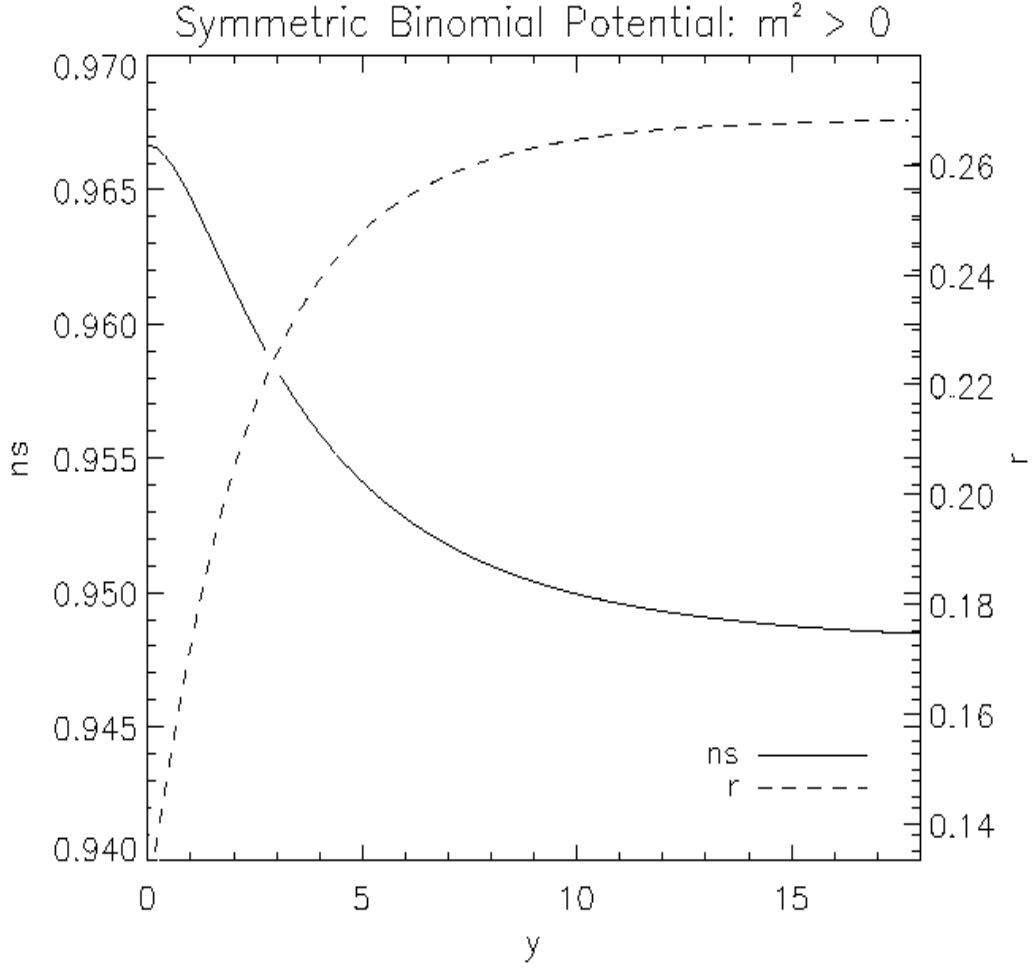,width=14cm,height=14cm}}
\caption{The scalar index $n_s$ and the ratio $r$ as functions of 
$ y =\kappa \, N $ for $m^2 > 0$  and $ N = 60 $
(chaotic inflation), with the binomial potential eq.(\ref{binonr}).
Both $n_s$ and $r$ monotonically interpolate
between their limiting values corresponding to the pure monomials 
$ \varphi^2 $ and $ \varphi^4 $.} \label{nsrm2p}
\end{figure}

\begin{figure}[htbp]
{\epsfig{file=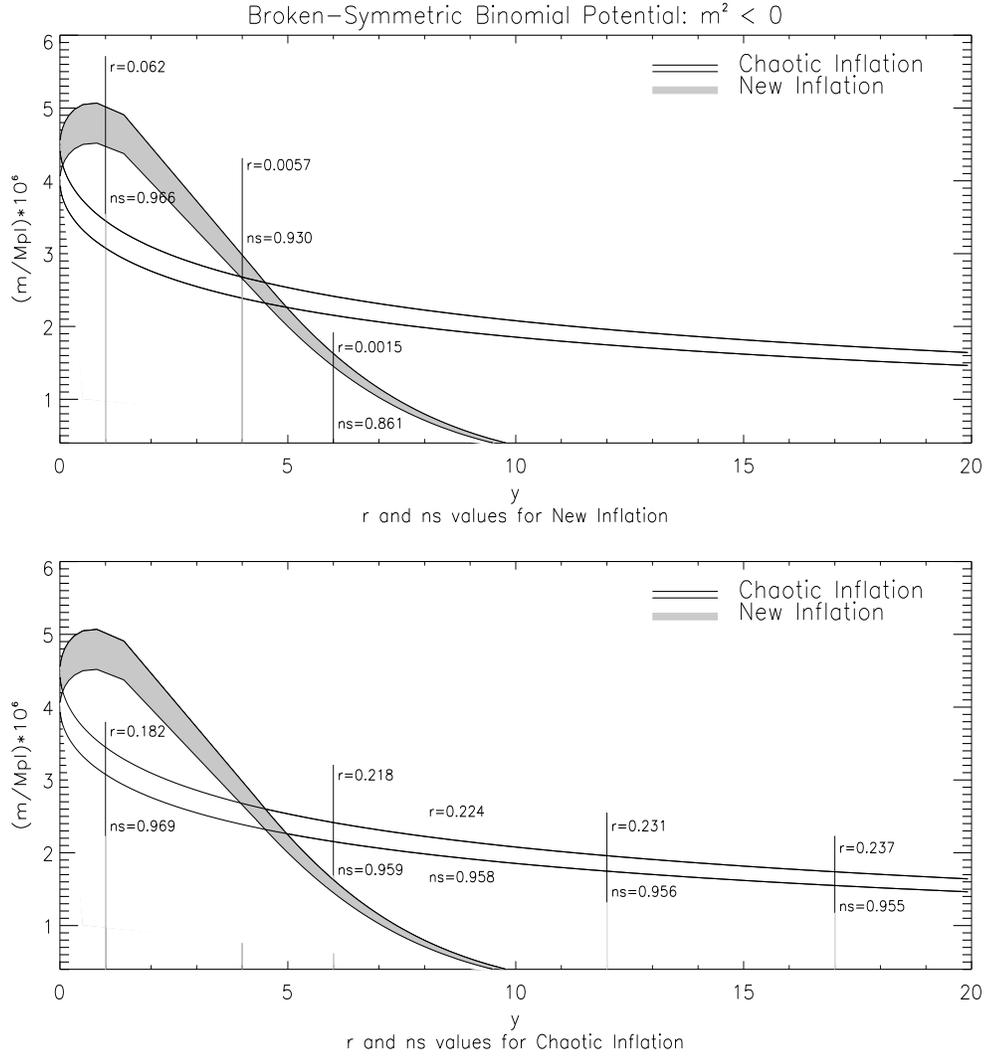,width=14cm,height=14cm}}
\caption{Upper and lower bounds for $10^6 \; m/M_{Pl}$ for $ N = 60 $ as a 
function of $ y = \kappa \, N$ with the binomial potential eq.(\ref{binor})
for $m^2 < 0$. Lower branches describe chaotic inflation and upper branches 
correspond to new inflation.} 
\label{xm2n}
\end{figure}

\begin{figure}[htbp]
{\epsfig{file=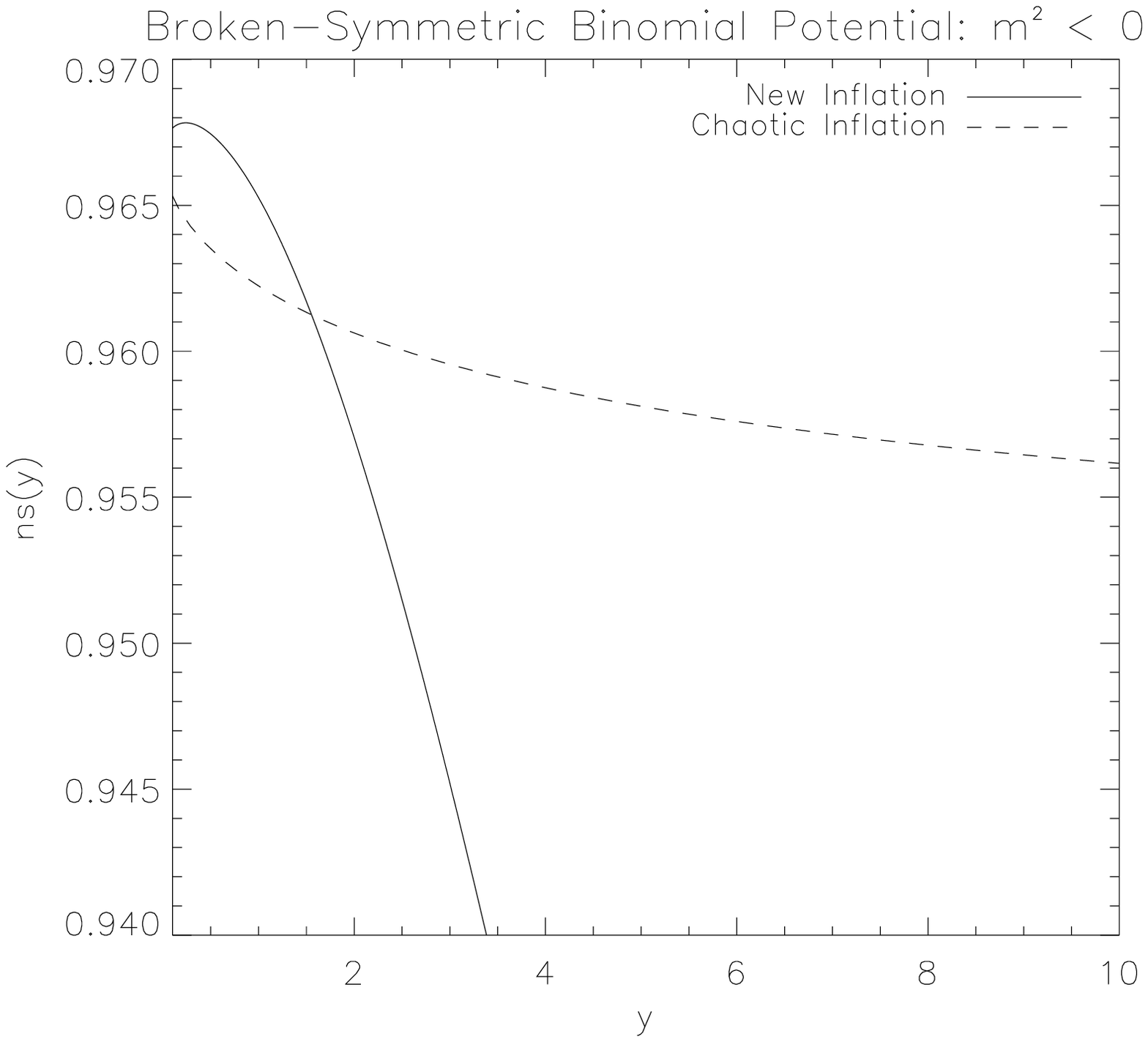,width=14cm,height=14cm}}
\caption{$n_s$ as a function of $ y = \kappa \, N $ for $m^2 < 0$ and
$ N = 60$, 
chaotic and new inflation, with the binomial potential eq.(\ref{binor}).} 
\label{nsm2n}
\end{figure}

\begin{figure}[htbp]
{\epsfig{file=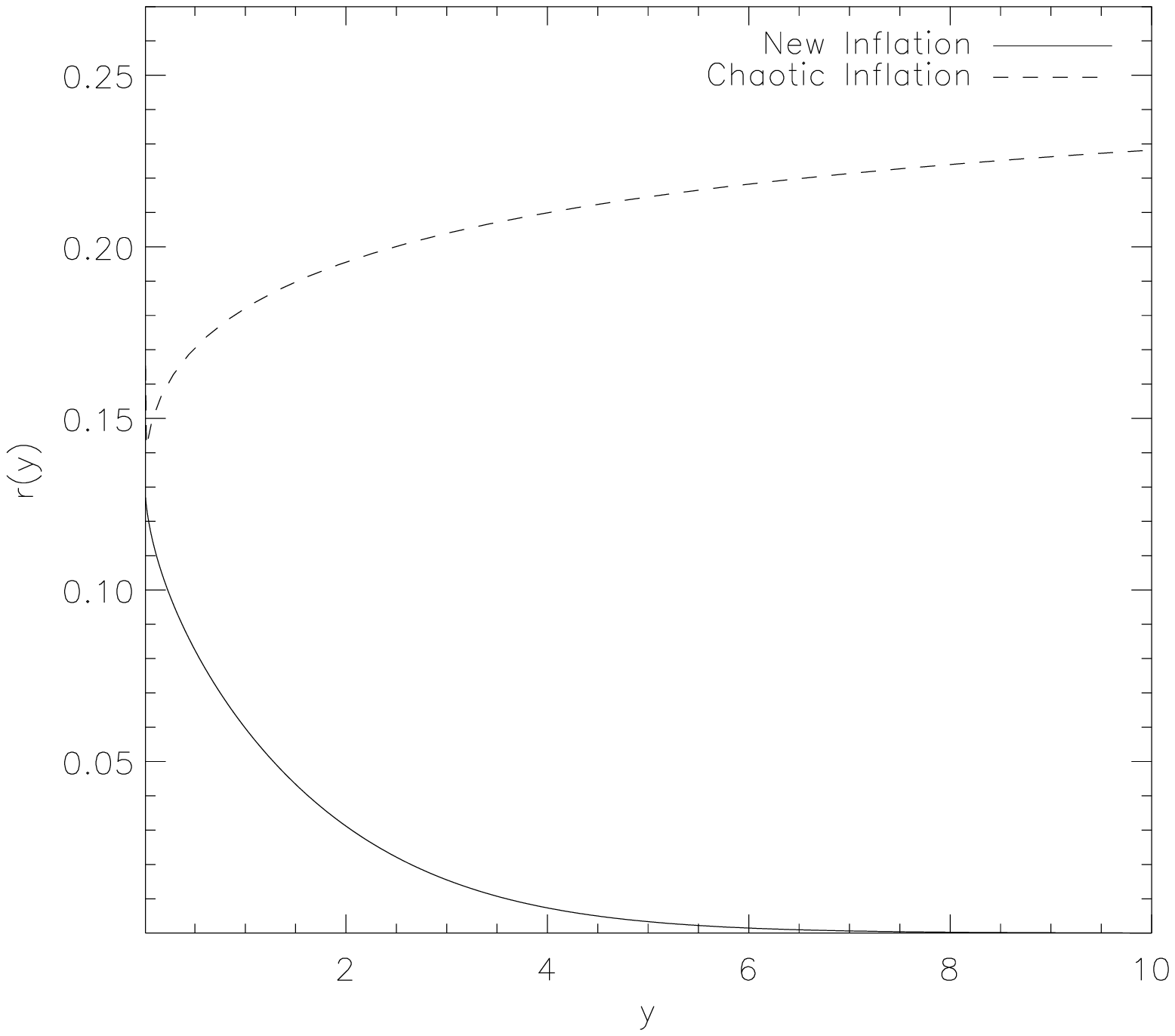,width=14cm,height=14cm}}
\caption{$r$ as a function of $  y = \kappa \, N $ for $m^2 < 0$ and
$ N = 60$, 
chaotic and new inflation, with the binomial potential eq.(\ref{binor}).} 
\label{rm2n}
\end{figure}

\begin{figure}[htbp]
{\epsfig{file=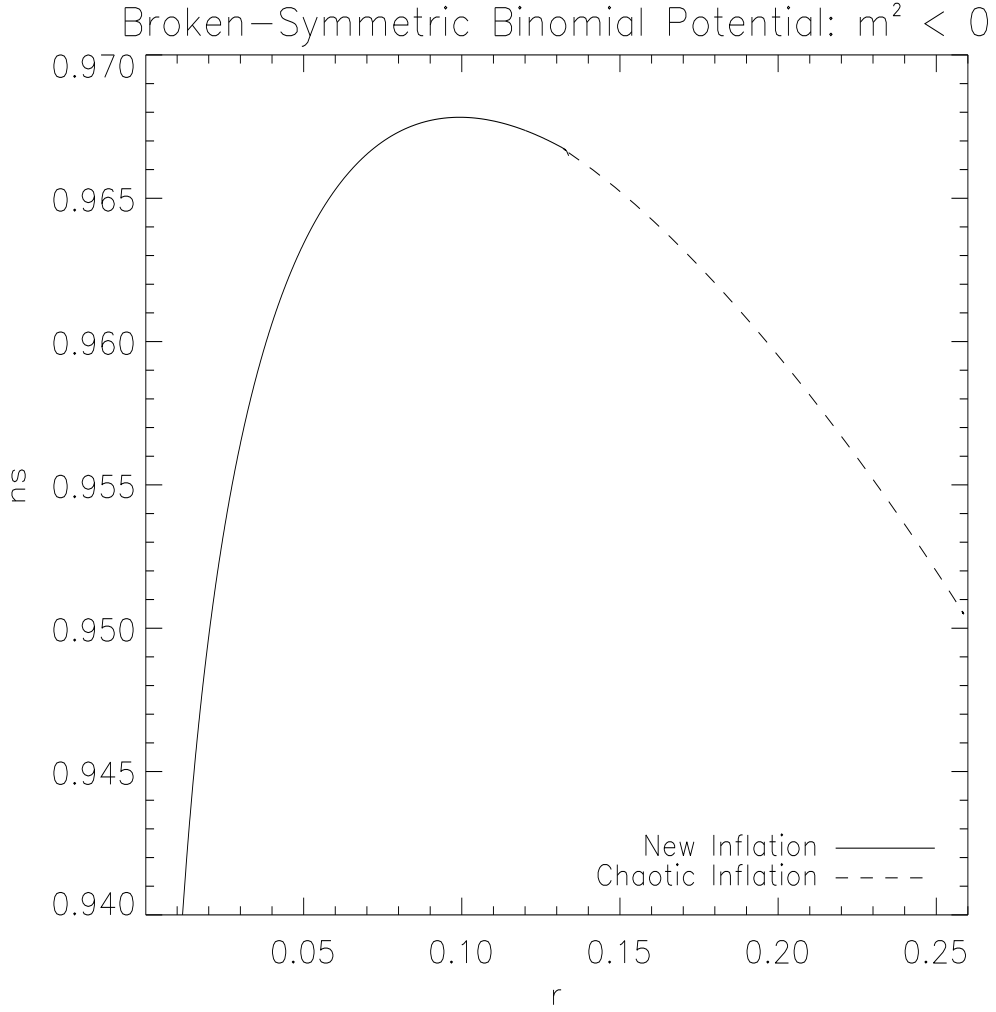,width=14cm,height=14cm}}
\caption{$n_s$ vs. $r$ for $m^2 < 0$, new and chaotic inflation,
with the binomial potential eq.(\ref{binor}).} \label{nsrm2n}
\end{figure}

\medskip

In the slow-roll approximation valid for $ {\dot \varphi} \ll \varphi $
we can approximate the number of efolds from the time $\tau$ till the end
of inflation as
\be \label{nef}
N(\tau) = \int_{\tau}^{\tau_0} h(\tau) \; d\tau = 
-3 \int_{\varphi(\tau)}^{\varphi_0}
\frac{h^2}{v'(\varphi)} \; d\varphi = -  \int_{\varphi(\tau)}^{\varphi_0}
\frac{v(\varphi)}{v'(\varphi)} \; d\varphi  \; .
\ee
where $ \varphi_0 $ is the inflaton field by the end of inflation.
That is, modes with co-moving wavenumber $ k = m \; h(\tau) \; a(\tau) $
cross the horizon for the first time at the time $ \tau , \; N(\tau)$ efolds
before the end of inflation.

\medskip

The spectral indices and $r$ can be expressed in terms of the slow roll parameters 
as\cite{hu},
\be\label{indesp}
n_s = 1 -6 \; \epsilon + 2 \, \eta \quad , \quad r = 16 \, \epsilon
\ee
where to dominant order in slow roll,
\be\label{epseta}
\epsilon =  \frac12 \,  M_{Pl}^2 \; \left( \frac{V'}{V} \right)^2 
= \frac12 \; \left( \frac{v'(\varphi)}{v(\varphi)} \right)^2 \quad , \quad 
\eta =   M_{Pl}^2 \;  \frac{V''}{V} = \frac{v''(\varphi)}{v(\varphi)} \; .
\ee
The amplitude of adiabatic perturbations is expressed as
\be \label{amplis}
|{\delta}_{k\;ad}^{(S)}|^2  =
\frac{1}{12 \, \pi^2 \;  M_{Pl}^6} \; \frac{V^3}{V'^2}= \frac{1}{12 \, \pi^2}
\frac{|m^2|}{M_{Pl}^2} \frac{v^3(\varphi)}{v'^2(\varphi)} \; .
\ee

\begin{figure}[htbp]
{\epsfig{file=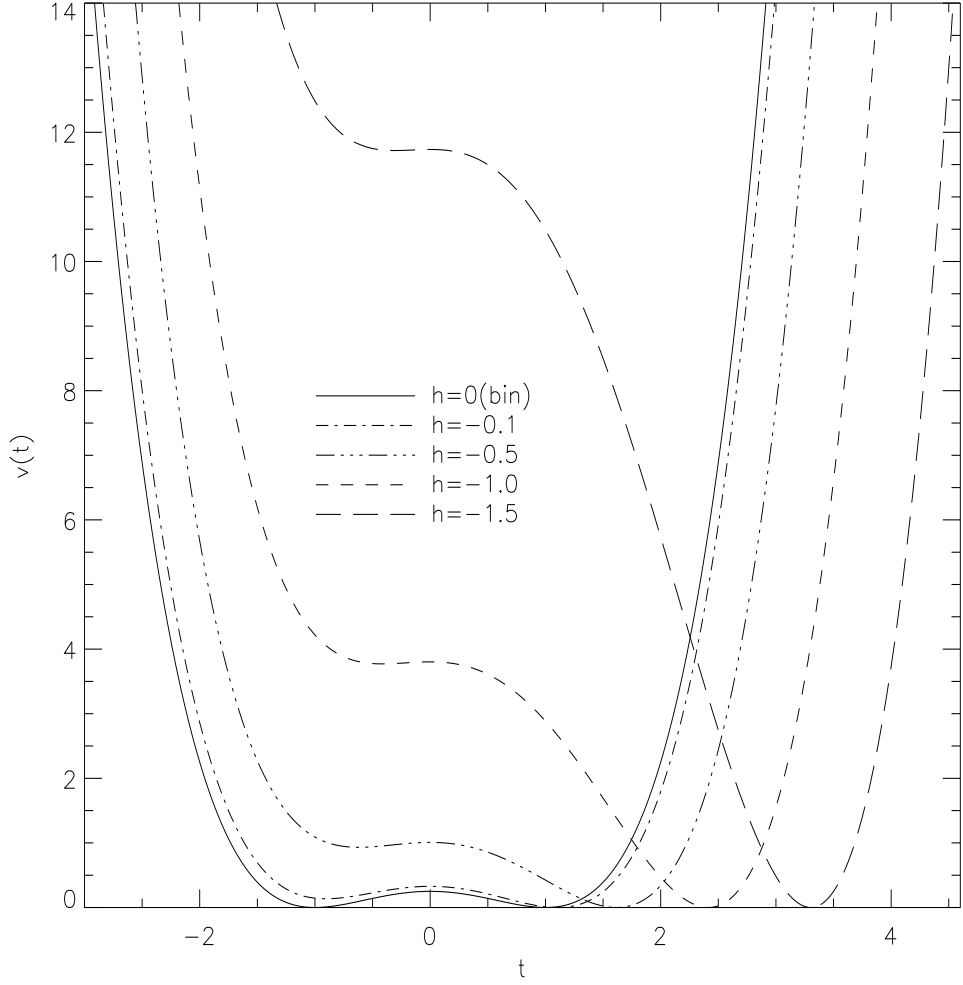,width=14cm,height=14cm}}
\caption{The inflaton trinomial potential $\kappa \, v(\varphi)$ 
eq.(\ref{trino}) vs. $ t=\sqrt{z} = \varphi \, \sqrt{\frac{\kappa}{8}} $ 
for several values of $ h < 0 $. The absolute minimum moves to the right for
growing $ |h| $ and the potential becomes more and more asymmetric.
$ h \equiv\gamma \; \sqrt{\frac{8}{\kappa}}  = 
\frac{g}{2 \; \sqrt{\lambda}} $.}
\label{potencial}
\end{figure}

\begin{figure}[htbp]
{\epsfig{file=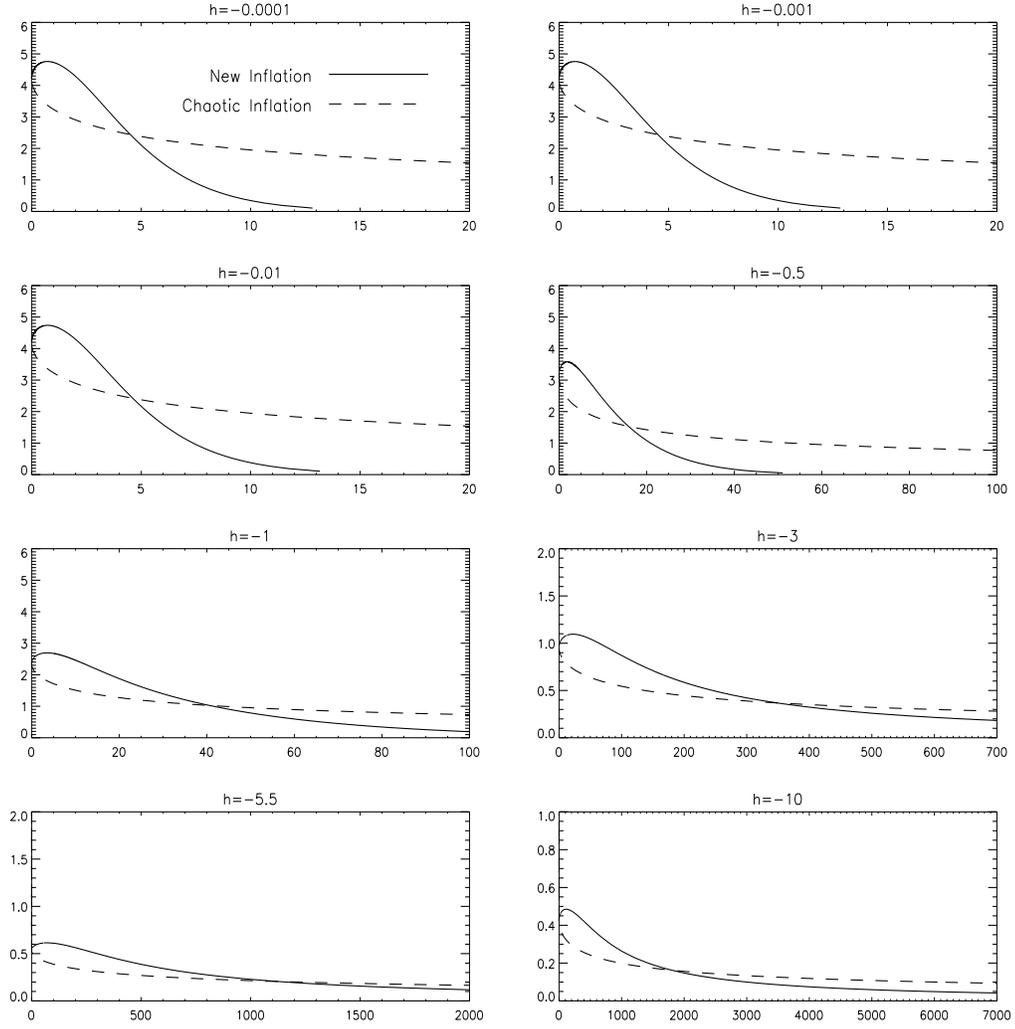,width=14cm,height=14cm}}
\caption{The central value for $ x = 10^6 \; \frac{m}{M_{Pl}} $ as a function
of $ y = \kappa \, N $ for different values of $ h < 0 $ and $ N = 60$
with the trinomial potential eq. (\ref{trino}) as given by 
eq.(\ref{cotxh}). The upper branches correspond to new inflation while the lower
 branches correspond to chaotic inflation (compare with fig. \ref{trinsmulti}).
Notice that the upper and lower bounds for $ x $ differ from the central value
in only $ \pm 6\% $.}
\label{trimasamulti}
\end{figure}

\medskip

As explained above, we can take  $ \varphi_0 = 0 $ for chaotic inflation 
while for new inflation $ \varphi_0 =\frac{1}{\sqrt\kappa}$. We obtain in this 
way inserting eqs.(\ref{binonr}) and (\ref{binor}) into eq.(\ref{nef}),
\be \label{znr}
\kappa \, N(\varphi) =  \frac{\kappa}{8} \, \varphi^2 + 
\log\left[ 1 + \frac{\kappa}{8} \, \varphi^2\right]\quad  m^2 > 0 
\ee
and
\be \label{zr}
\kappa \, N(\varphi) =   \frac{\kappa}{8} \, \varphi^2 -1- 
\log\left[ \frac{\kappa}{8} \, \varphi^2\right] \quad  m^2 < 0 \; .
\ee
Eqs.(\ref{znr})-(\ref{zr}) can be written in an unified form by introducing 
the new variables $z_{ub}$ and $z_{b}$
\be\label{defz}
z_{ub} \equiv  1 +  \frac{\kappa}{8} \, \varphi^2 \quad , 
\quad z_{b} \equiv  \frac{\kappa}{8}
 \, \varphi^2 \quad , \quad \kappa = 8 \, \lambda \; 
\frac{M_{Pl}^2}{|m^2|} \; .
\ee  
That is,
\be \label{kNz}
 \kappa \, N(\varphi) = z_{ub} -1 + \log z_{ub} \quad  m^2 > 0 \quad , \quad
 \kappa \, N(\varphi) =  z_b -1 - \log z_b \quad  m^2 < 0 \; ,
\ee
and we have
\bea
&&1 < z_{ub} < \infty \quad , \quad \mbox{chaotic~inflation} 
\quad m^2 > 0 \; , \cr \cr
&& 0 < z_b < 1\quad , \quad \mbox{new~inflation} \quad m^2 < 0 \; ,\cr \cr
&& 1  < z_b < \infty \quad , \quad \mbox{chaotic~inflation}  
\quad m^2 < 0 \ . \nonumber
\eea
Using eqs.(\ref{epseta}) 
$ n_s $ and $r$ can be expressed in terms of $z_{ub}$ and 
$z_{b}$ 
\bea\label{indicao}
&& m^2 > 0 \quad : \quad
n_s = 1 - \kappa \; \frac{3 \, z_{ub}^2 - z_{ub} + 
2}{(z_{ub}-1)(z_{ub}+1)^2} \quad , \quad
r = 16 \, \kappa \; \frac{z_{ub}^2}{(z_{ub}-1)(z_{ub}+1)^2} \; ; 
\\ \cr
&&  m^2 < 0 \quad : \quad
n_s = 1 - \kappa \; \frac{3 \, z_b + 1}{(z_b-1)^2}  \quad , \quad
r = 16 \, \kappa \; \frac{z_b}{(z_b-1)^2} \; ,
\label{indinue} \; 
\eea
and the  amplitudes of adiabatic perturbations eq.(\ref{amplis}) read
\bea\label{deltcao}
&&|{\delta}_{k\;ad}^{(S)}|^2  = 
\frac1{12 \, \pi^2} 
\frac{m^2}{M_{Pl}^2}\frac{(z_{ub}-1)^2(z_{ub}+1)^3}{\kappa^2
\; z_{ub}^2 } \quad m^2 > 0 \; , \\ \cr
&&|{\delta}_{k\;ad}^{(S)}|^2  = \frac1{12 \, \pi^2} 
\frac{|m^2|}{M_{Pl}^2}\frac{(z_b-1)^4}{\kappa^2 \; z_b} \quad 
\quad m^2 < 0  \; . \label{deltnue}
\eea
The variables $z_{ub}$ and $z_{b}$ are functions of $ \kappa $ times the 
number of efolds $ N $ defined by eqs.(\ref{znr})-(\ref{zr}). Hence, 
eqs.(\ref{kNz})-(\ref{indinue}) provide the spectral 
indices as functions of $ \kappa \; N $ in a parametric way. 

Eqs.(\ref{deltcao})-(\ref{deltnue}) permit to express the mass ratio 
$ \frac{|m^2|}{M_{Pl}^2} $ in terms of the 
amplitude of adiabatic perturbations, and hence determine 
$ \frac{|m^2|}{M_{Pl}^2} $ using the WMAP values $\Delta_{\pm} $ 
\cite{WMAP} for $ |{\delta}_{k\;ad}^{(S)}|^2 $,
\be\label{cotad}
\Delta_- \equiv 0.194 < 10^8 \; |{\delta}_{k\;ad}^{(S)}|^2 < 
\Delta_+ \equiv 0.244 \; .
\ee
In terms of the variable,
\be\label{x}
x \equiv 10^6 \; \frac{m}{M_{Pl}} \; ,
\ee
the observational bounds eq.(\ref{cotad}) imply
\be\label{cotx}
\frac{200 \, \sqrt3 \, \pi}{N} g(z) \; \sqrt{\Delta_-} < x 
< \frac{200 \, \sqrt3 \, \pi}{N} g(z) \; \sqrt{\Delta_+} \; ,
\ee
where
\be  
 m^2 > 0\quad : \quad
g_{ub}(z_{ub}) = \frac{\kappa \, N \, z_{ub}}{(z_{ub}-1)(z_{ub}+1)^{\frac32}}
\quad ; \quad m^2 < 0 \quad : \quad
g_b(z_b) = \frac{\kappa \, \sqrt{z_b} \; N}{(z_b-1)^2} \; .
\ee
Recall that $ \kappa \, N $ is a function of 
$ z_{ub} $ or $ z_b $, respectively, according to eq.(\ref{kNz}). 
Notice that the upper and lower bounds $ \sqrt{\Delta_{\pm}} $ are quite 
close\cite{WMAP}:
\be\label{ampmap}
10^4 \;  |{\delta}_{k\;ad}^{(S)}|=\sqrt{\Delta_{\pm}} = 0.467 \pm 0.027 \; ,
\ee
which corresponds to $  \pm 6 \% $. 

\medskip

For $ m^2> 0 $ (chaotic inflation) we plot in fig.  \ref{xm2p} the upper and 
lower bounds for $ x =  10^6 \; \frac{m}{M_{Pl}} $ given in eqs.(\ref{cotx}) 
as functions of $ \kappa \, N $,
and in fig. \ref{nsrm2p} we plot $n_s$ and $r$ as 
functions of $ \kappa \, N $. We see that both $n_s$ and $r$ for
$ 0 < \kappa \, N < \infty $ monotonically interpolate between their
 limiting values given by eqs.(\ref{monofi2}) and (\ref{monofi4})
corresponding to the purely quadratic and quartic potentials, respectively.
We used $ N = 60 $ in the plots. 
Very similar results are obtained for $ N = 50 $.

In fig.  \ref{xm2n} we plot the upper and lower bounds for 
$x =  10^6 \; \frac{m}{M_{Pl}} $ given in eqs.(\ref{cotx}) as functions of 
$ \kappa \, N $ for $ m^2 < 0 $.
Both inflationary scenarios are displayed: new inflation for $ 0 < z_b < 1 $
and chaotic inflation for  $ 1  < z_b < \infty $. 

In figs. \ref{nsm2n} and \ref{rm2n} we plot $n_s$ and $r$ as functions of 
$ \kappa \, N $ for $ m^2 < 0 $, respectively. We see that $n_s$ and $r$ 
are two-valued functions of
$ \kappa \, N $. One branch corresponds to new inflation and the other 
branch to chaotic inflation.
$ \kappa \, N = 0 $ is a branch point where we recover the results for 
the purely (monomial) quadratic
potential  eq.(\ref{monofi2}). The result for the purely quartic potential 
eq.(\ref{monofi4})
is obtained from the chaotic inflation branch of  $n_s$ and $r$ in the
 $ \kappa \, N \to \infty $ limit.

We depict in fig. \ref{nsrm2n}  $n_s$ vs. $r$ for $ m^2 < 0 $ both for
new and chaotic inflation.

\subsubsection{\bf Limiting Cases}

Let us now consider limiting cases, first for unbroken symmetry ($m^2 > 0$) 
and then for broken symmetry ($m^2 < 0$).

\medskip

The limiting case  $ \kappa \to 0 $ corresponds to a vanishing quartic 
coupling [see eq.(\ref{gaka})]. That is, the inflaton potential reduces in 
this case to the quadratic piece $ v(\varphi) \to \frac12 \; \varphi^2$ for 
$m^2 > 0$.

For chaotic inflation in the $ \kappa \to 0 $ limit we obtain from  
eqs.(\ref{defz})-(\ref{indicao}),
\bea \label{monofi2}
&& z_{ub} \buildrel{\kappa\to 0}\over= 1 \quad  ,  \quad 
 n_s \buildrel{\kappa\to 0}\over= 1 - \frac{2}{N} \simeq 0.96 \quad , \quad
r \buildrel{\kappa\to 0}\over= \frac{8}{N}\simeq 0.16 \;  , \cr \cr
&&\mbox{chaotic~inflation,~} \; v(\varphi) = \frac12 \, \varphi^2 \quad  ,  
\quad  |{\delta}_{k\;ad}^{(S)}|^2 \buildrel{\kappa\to 0}\over= 
\frac1{6 \, \pi^2} \frac{N^2 \, m^2}{M_{Pl}^2} \quad  , \quad
g_{ub}(1) = \frac1{\sqrt2} \; .
\eea

\medskip

The opposite limit $ \kappa \to \infty $ corresponds to a vanishing mass 
$ m^2 $ [see eq.(\ref{gaka})]. That is, the inflaton potential becomes in this 
case purely quartic $ V(\phi) \to \frac{\lambda}{4}\; \phi^4 $.

For chaotic inflation in the $ \kappa \to \infty $ limit we obtain from 
eqs.(\ref{defz})-(\ref{indicao})
\bea\label{monofi4}
&& z_{ub} \buildrel{\kappa\to +\infty}\over= +\infty \quad  ,  \quad 
n_s \buildrel{\kappa\to +\infty}\over= 1 - \frac{3}{N}  \simeq 0.94
\quad ,  \cr \cr
&&r \buildrel{\kappa\to +\infty}\over= \frac{16}{N}\simeq 0.32 \quad , \quad
\mbox{chaotic~inflation,~purely~quartic~}V(\phi) \cr \cr
&&|{\delta}_{k\;ad}^{(S)}|^2 \buildrel{\kappa\to +\infty}\over= 
\frac{2}{3 \, \pi^2} \frac{N^3 \, m^2}{M_{Pl}^2} \quad  , \quad
g_{ub}(z) \buildrel{\kappa\gg 1}\over= \frac{1}{\sqrt{\kappa \, N}}
 \; .
\eea 
For $m^2 < 0$ (i. e. broken symmetry) and  $ \kappa \to 0 $ we find from  
eq.(\ref{zr}) that 
$ z_b \to 1 $ and then $ n_s , \;  r $ and $ |{\delta}_{k\;ad}^{(S)}|^2 $
in eqs.(\ref{indinue}) and (\ref{deltnue}) tend to the 
same values than for the unbroken symmetry case eq.(\ref{monofi2}).

\medskip  

The limiting case  $ \kappa \to \infty $ for broken symmetry and chaotic
inflation leads  from eq.(\ref{kNz}) to $ z_b \to +\infty $ and we recover the 
same values as in eq.(\ref{monofi4}) after using  eqs.(\ref{indinue}) and 
(\ref{deltnue}).

\medskip

\begin{figure}[htbp]
{\epsfig{file=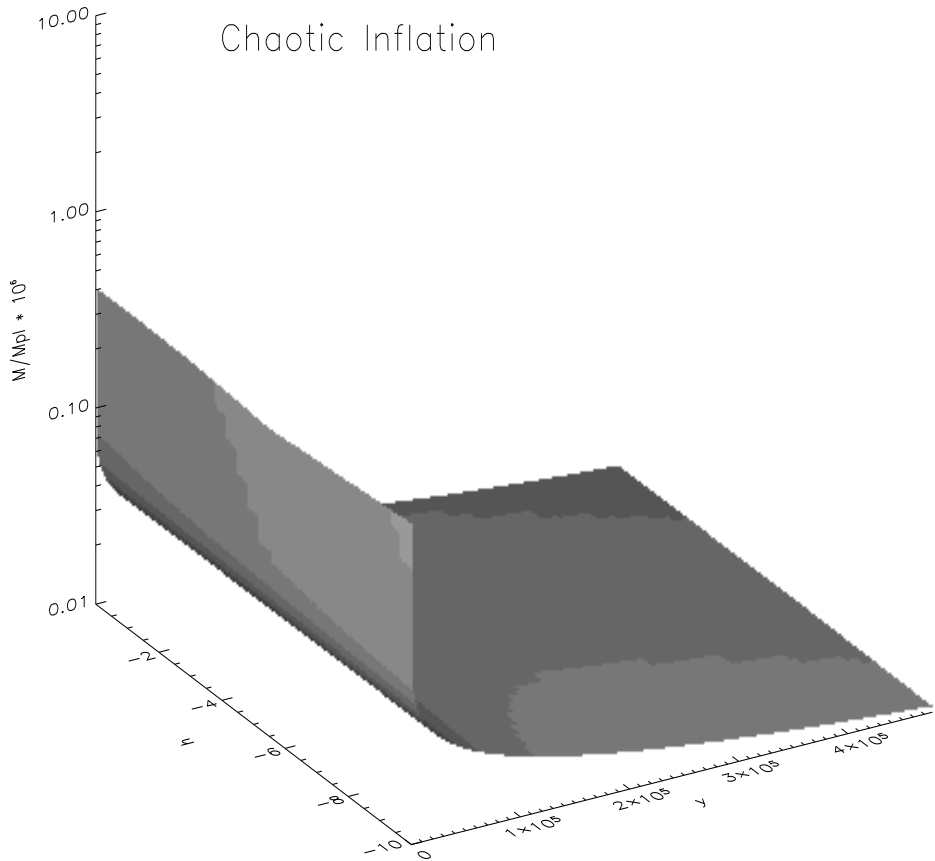,width=14cm,height=14cm}}
\caption{The central value for $ x = 10^6 \; \frac{m}{M_{Pl}} $ as a function
of $ y = \kappa \, N $ for $ h < 0 $ and $ N = 60 $ with the trinomial 
potential eq. (\ref{trino}) as given by eq.(\ref{cotxh}) for chaotic inflation. 
Notice that the upper and lower bounds 
for $ x $ differ from the central value in only $ \pm 6\% $.}
\label{trimasasurfCAO}
\end{figure}

\begin{figure}[htbp]
{\epsfig{file=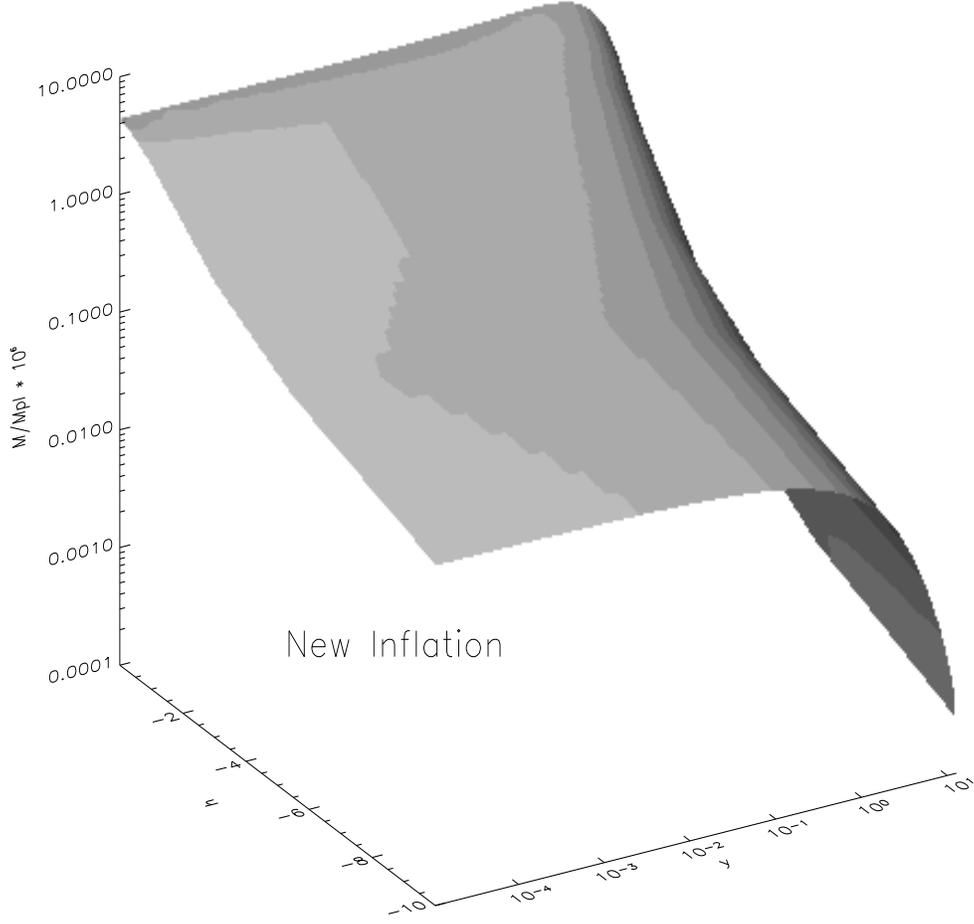,width=14cm,height=14cm}}
\caption{The central value for $ x = 10^6 \; \frac{m}{M_{Pl}} $ as a function
of $ y = \kappa \, N $ for $ N = 60$ and $ h < 0 $ with the trinomial potential 
eq. (\ref{trino})
as given by eq.(\ref{cotxh}) for new inflation.}
\label{trimasasurfNEW}
\end{figure}

\begin{figure}[htbp]
{\epsfig{file=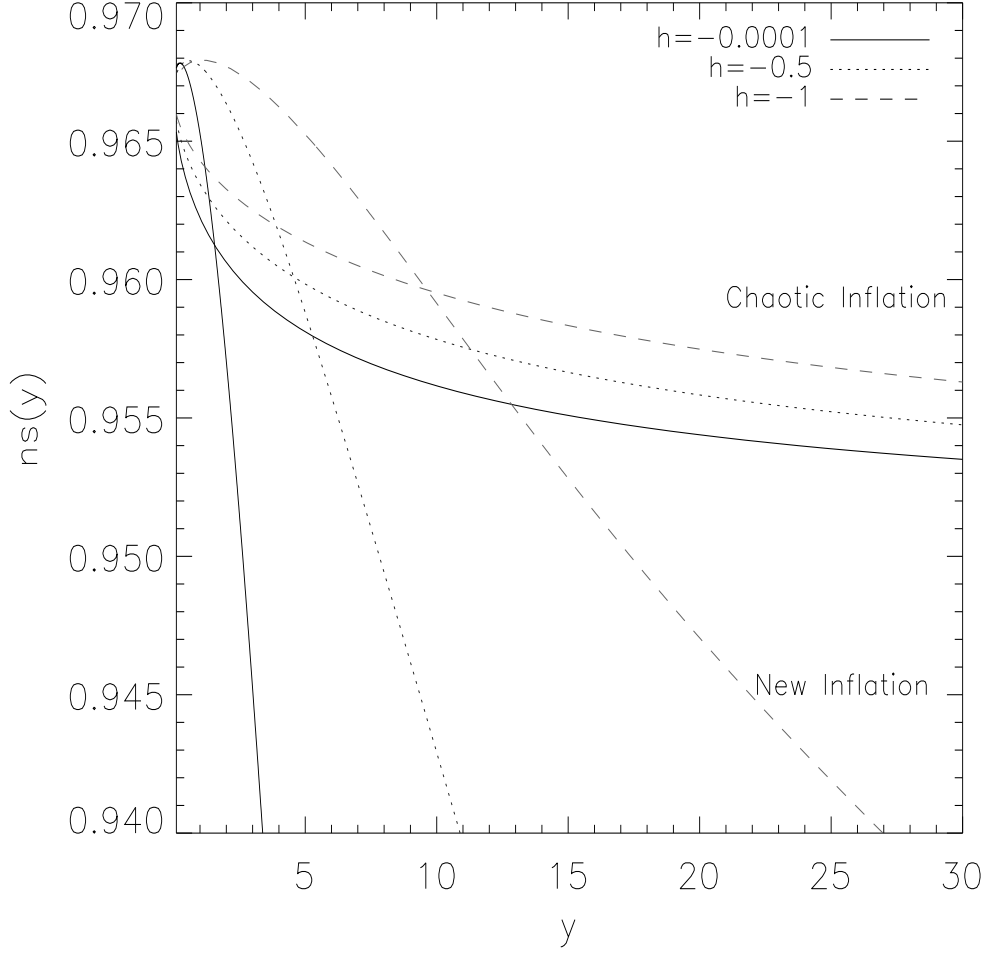,width=14cm,height=14cm}}
\caption{ $n_s$ as a function of $ y = \kappa \, N $ for $ N = 60$ and 
different values of $ h < 0 $ with the trinomial potential eq. (\ref{trino})
as given by eqs.(\ref{ntrino})-(\ref{nstrino}).
The upper branches correspond to new inflation while the lower
 branches correspond to chaotic inflation.}
\label{trinsmulti}
\end{figure}

\begin{figure}[htbp]
{\epsfig{file=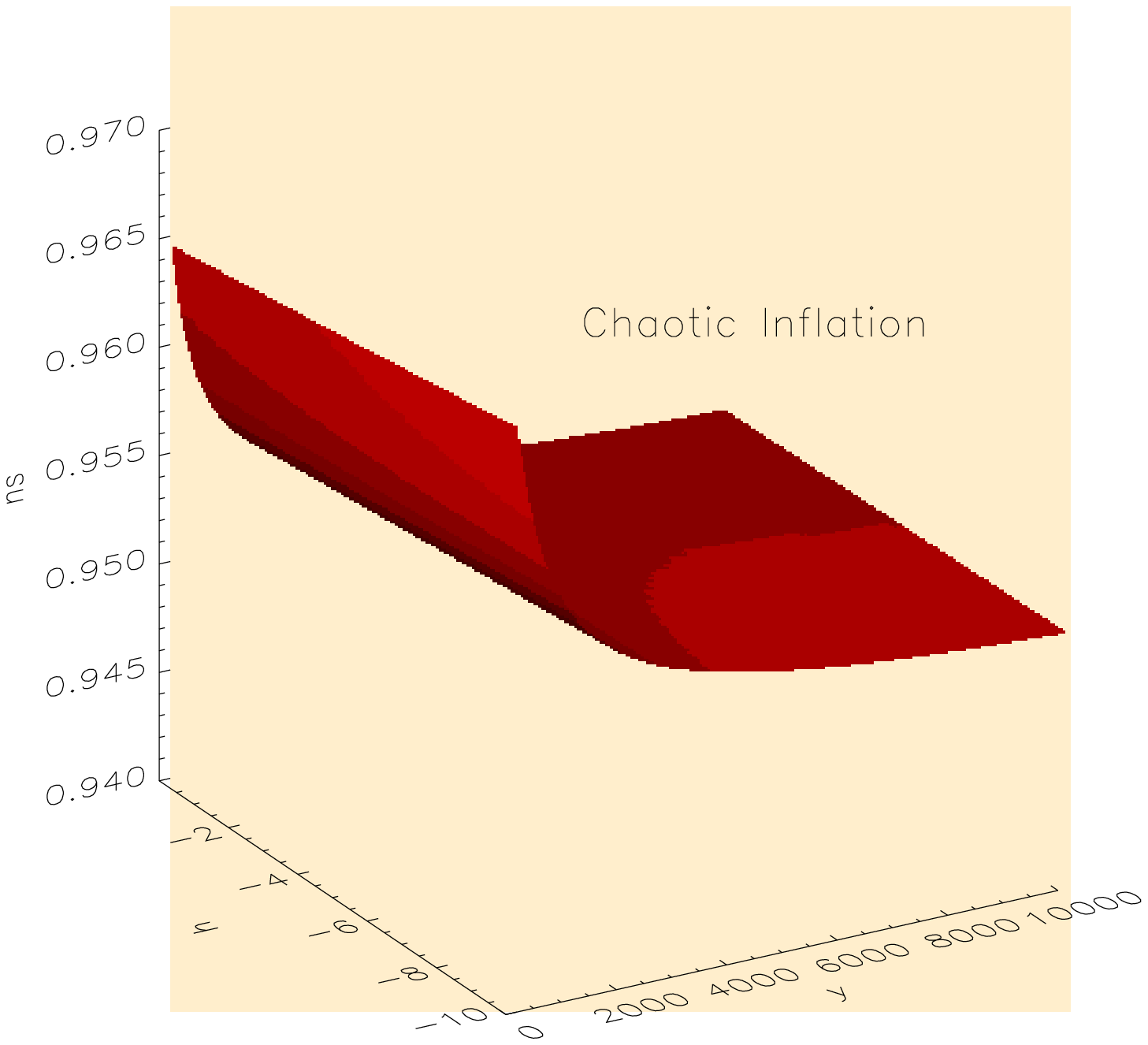,width=14cm,height=14cm}}
\caption{ $n_s$ as a function of $ y = \kappa \, N $ for $ N = 60$ and 
$ h < 0 $ with the trinomial potential eq. (\ref{trino})
as given by eqs.(\ref{ntrino})-(\ref{nstrino}) for chaotic inflation.}
\label{trinssurfCAO}
\end{figure}

\begin{figure}[htbp]
{\epsfig{file=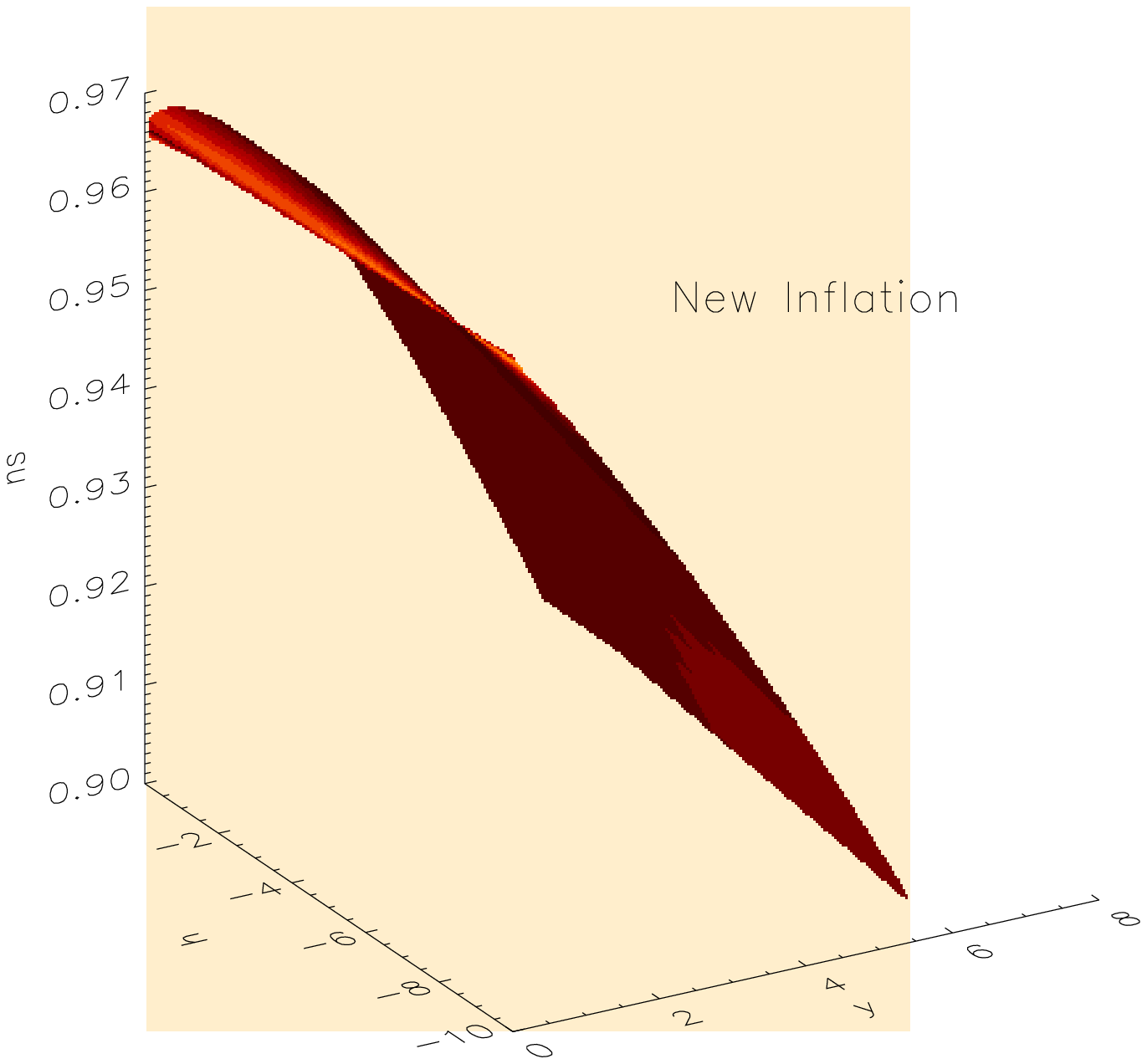,width=14cm,height=14cm}}
\caption{ $n_s$ as a function of $ y = \kappa \, N $  for $ N = 60$ and 
$ h < 0 $ with the trinomial potential eq. (\ref{trino})
as given by eqs.(\ref{ntrino})-(\ref{nstrino}) for new inflation.}
\label{trinssurfNEW}
\end{figure}

When  $ \kappa \to \infty $ for broken symmetry and new inflation we
have from eq.(\ref{kNz}),
$$
z_b \buildrel{\kappa\gg 1}\over= e^{- \kappa \, N - 1} \; .
$$
This leads using  eqs.(\ref{indinue}) and (\ref{deltnue}) to
\bea\label{nsrpuro4}
&&n_s \buildrel{\kappa\gg 1}\over= 1 - \kappa \quad  , \quad 
r \buildrel{\kappa\gg 1}\over= 16 \,  \kappa \, 
e^{- \kappa \, N - 1} \quad , \quad
\mbox{new~inflation,~} V(\phi)\buildrel{\kappa\gg 1}\over
= \frac{\lambda}{4} \; \phi^4\mbox{~in~the~limit}\; , \cr \cr
&& 
|{\delta}_{k\;ad}^{(S)}|^2  \buildrel{\kappa\gg 1}\over= \frac1{768 \, \pi^2} 
\frac{m^4}{\lambda \; M_{Pl}^4} \;  e^{\kappa \, N + 1} 
 \quad , \quad g_b(z) \buildrel{\kappa\gg 1}\over= \kappa \, N \;
 e^{- \frac12 \,\kappa \, N - \frac12} \to 0 \; .
\eea
In this limit, new inflation yields a very small ratio $r$ together
with an index $n_s$ well below unit, while the bound on the inflaton mass 
eq.(\ref{cotx}) becomes very small as compared with $ M_{Pl} $ since 
$ g_b(z) $ decreases exponentially with $\kappa$. 

\medskip

We see from fig. \ref{nsrm2p} that for chaotic inflation and
both signs of $ m^2 $,
$$
 n_s \leq 1 - \frac{2}{N} \simeq 0.96 \qquad  \mbox{chaotic~inflation,}
$$ 
Fig. \ref{nsm2n}
shows that $n_s$ for new inflation has a maximum at $ y = \kappa N =
0.2386517\ldots $ and then 
$$
n_s \leq 1 - \frac{1.558005\ldots}{N} \simeq 0.9688 \qquad  
\mbox{new~inflation}.
$$ 
Figs. \ref{nsrm2p} and \ref{rm2n} show that for chaotic inflation and 
both signs of $ m^2 $,
$$ 
 0.16 \simeq \frac{8}{N} < r < \frac{16}{N} \simeq 0.32 \qquad 
\mbox{chaotic~inflation,}
$$
while for new inflation ($ m^2 < 0 $),
$$ 
0 < r < \frac{8}{N} \simeq 0.16 \qquad \mbox{new~inflation.}
$$ 
Notice that new inflation can give {\bf small values} for $r$ 
with a $ n_s $ significantly  {\bf below unit} and a low value for 
$|m|^2/M_{Pl}^2$ while $r$ in chaotic inflation is bounded from below by 
$ \frac{8}{N} \simeq 0.15 $. We  come back to this point when analyzing 
the trinomial inflaton potential in sec. IIIB below.

\subsection{The trinomial inflaton potential}

We consider in this section inflation arising from the trinomial
inflaton potential with negative squared mass
\be\label{trino}
V(\phi) = |m|^2 \;  M_{Pl}^2 \; v(\varphi) \quad \mbox{where} \quad
v(\varphi) = -\frac12 \; \varphi^2 + \frac23 \; \gamma \; 
\varphi^3 + \frac1{32} \; \kappa  \; \varphi^4 + v_0 \; .
\ee
This potential has three extremes: a local maximum at $ \varphi = 0 $ and two
local minima at $ \varphi = \varphi_{\pm} $ where
\be
\varphi_+ = \frac{8}{\kappa} \left[
\sqrt{\gamma^2 + \frac{\kappa}{8}} - \gamma\right]
\quad , \quad \varphi_- =-\frac{8}{\kappa}\left[\sqrt{\gamma^2 + 
\frac{\kappa}{8}} + \gamma\right] \; .
\ee
The absolute minimum of  $ v(\varphi) $ is at $ \varphi =\varphi_- $ for 
$ \gamma > 0 $ and at $ \varphi =\varphi_+ $ for $ \gamma < 0 $.
We choose $ v_0 $ such that $ v(\varphi) $ vanishes at its absolute minimum.
Such condition gives,
\be\label{defw}
v_0 =\frac{2}{\kappa} \; w(h) \quad \mbox{where}  \quad 
 w(h) \equiv \frac83 \, h^4 + 4 \, h^2 + 1 + \frac83 \, |h| \, \Delta^3 
\quad , \quad h \equiv \gamma \; \sqrt{\frac{8}{\kappa}} 
\quad \mbox{and} \quad \Delta \equiv \sqrt{h^2 + 1}\; .
\ee
The parameter $ h $ reflects how asymmetric is the potential.
Notice that  $ v(\varphi) $ is invariant under the changes
$ \varphi \to - \varphi , \;  \gamma \to - \gamma $. We can hence restrict
ourselves to a given sign for $ \gamma $. Without loss of 
generality, we choose $ \gamma < 0 $ and shall
work with positive fields $ \varphi $.

We plot in fig. \ref{potencial} the potential $ v(\varphi) $ times $ \kappa $ 
as a function of $ t = \sqrt{z_b} = 
\varphi \, \sqrt{\frac{\kappa}{8}} $ for several values of $ h < 0 $.
For growing $ |h| $  the potential becomes more asymmetric and 
its absolute minimum at $ \sqrt{z_+} = \sqrt{\frac{\kappa}{8}} \; 
\varphi_+ = \Delta + |h| $ moves to the right. 

\medskip

As for the binomial inflaton potential with $ m^2 < 0 $ we can have here new 
or  chaotic inflation depending on whether the initial field $ \varphi $ is in 
the interval $(0,\varphi_+)$ or in the interval $ (\varphi_+, +\infty) $, 
respectively. In both cases the number of efolds follows 
from eqs.(\ref{nef}) and (\ref{trino})
where the field by the end of inflation is $ \varphi_0 \simeq \varphi_+ $. 
We thus find for the number of efolds between the time $ \tau $ and the end 
of inflation,
\bea\label{ntrino}
&&\kappa \, N(z_b) = z_b - 2 \; h^2 -1 - 2  \; |h|  \; \Delta + \frac43 \; 
|h|  \; \left( |h| + \Delta - \sqrt{z_b} \right) + \\ \cr
&&+\frac{16}{3} \; |h| \;  (\Delta + |h| ) \; \Delta^2  \; 
\log\left[\frac12 \left(1 +  \frac{\sqrt{z_b} -  |h|}{\Delta}\right)\right] - 
2 \, w(h) \, \log\left[\sqrt{z_b} \; (\Delta - |h|)\right] \; , \nonumber
\eea
where $ z_b = \frac{\kappa}{8} \; \varphi^2 $  is defined by eq.(\ref{defz})
and $ w(h) $ is given by eq.(\ref{defw}). In the variable $z$ we have new 
inflation for $ 0 < z < z_+ $ and chaotic inflation for $  z_+ < z < +\infty $.

$\kappa \, N(z_b)$ in eq.(\ref{ntrino}) as a function of $ z_b $ has
its minimum at $ z_b = z_+ \equiv
\frac{\kappa}{8} \; \varphi_+^2 = 2\, h^2 + 1 + 2 \, |h| \, \Delta $.
This corresponds to $ \sqrt{z_+} = \Delta + |h| $. When  $ \sqrt{z_b} \to 
\sqrt{z_+} , \; \kappa \, N(z_b) $ vanishes quadratically,
$$
\kappa \, N(z_b) \buildrel{z_b \to z_+}\over= 2 \; 
\left(\sqrt{z_b} - \sqrt{z_+}\right)^2 + {\cal O} 
\left(\left[\sqrt{z_b} - \sqrt{z_+}\right]^3\right) \; .
$$
In the symmetric potential limit $ h \to 0 $, eq.(\ref{ntrino}) 
reduces to eq.(\ref{kNz}) as it must be. 

\bigskip

We obtain in analogous way from eqs. (\ref{epseta}) and (\ref{amplis}) the 
spectral indices, $r$ and the amplitude of adiabatic perturbations,
\bea\label{nstrino}
&&n_s=1 - 6 \,  \kappa \, \frac{z_b \; (z_b + 2  \, h  \, 
\sqrt{z_b} -1)^2}{\left[w(h)  -2 \, z_b+ \frac83  \, h  \,  z_b^{3/2} +
z_b^2\right]^2} + \kappa \, \frac{ 3 \, z_b+ 4 \, h  \, \sqrt{z_b} -1}{
w(h)- 2 \, z_b+ \frac83  \, h  \,  z_b^{3/2} + z_b^2} \quad , \\ \cr\cr
\label{rtrino}
&& r = 16 \,  \kappa \, \frac{z_b \; (z_b + 2  \, h  \, \sqrt{z_b} -1
)^2}{\left[w(h)  -2 \, z_b+ \frac83  \, h  \,  z_b^{3/2} +z_b^2
\right]^2}  \quad  , \\ \cr\cr
&&|{\delta}_{k\;ad}^{(S)}|^2  = \frac{1}{12 \, \pi^2} \frac{|m^2|}{M_{Pl}^2 
\;\kappa^2 }\frac{\left[w(h)- 2 \, z_b+ \frac83  \, h  \,  z_b^{3/2} + 
z_b^2\right]^3}{z_b \; (z_b + 2  \, h  \, \sqrt{z_b} -1)^2} \; .\label{dtrino}
\eea
The variable $z_{b}$ is a function of $ \kappa $ times the number 
of efolds $ N $ as defined by eq.(\ref{ntrino}). Hence, 
eqs.(\ref{nstrino})-(\ref{dtrino}) provide the spectral 
indices as functions of $ \kappa \; N $. 
In the $ h  \to 0 $ limit 
these equations reduce to eqs. (\ref{indinue}) and (\ref{deltnue}) 
for the binomial potential as it must be.

\begin{figure}[htbp]
{\epsfig{file=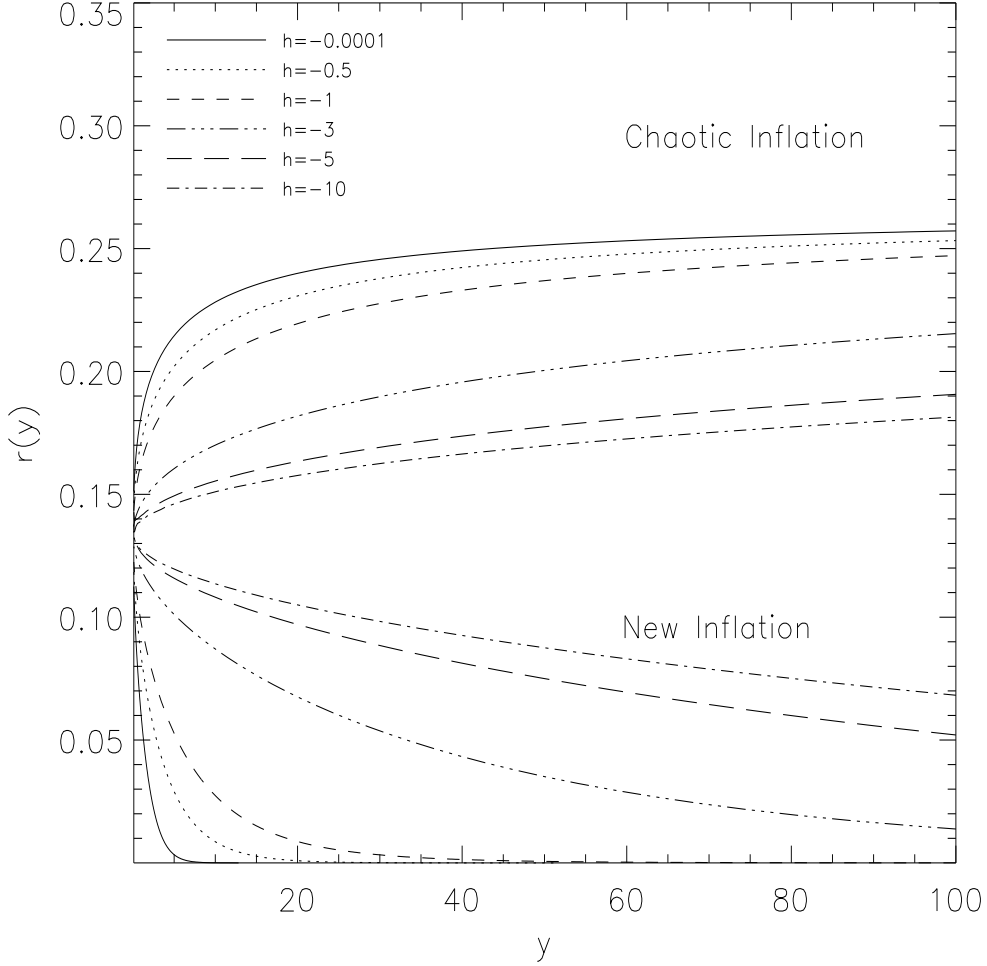,width=14cm,height=14cm}}
\caption{ $r$ as a function of $ y = \kappa \, N $  for $ N = 60$ and different 
values of $ h < 0 $ with the trinomial potential eq. (\ref{trino})
as given by eqs.(\ref{ntrino}), (\ref{ntrino}) and 
(\ref{rtrino}). The upper branches correspond to chaotic inflation while the 
lower branches correspond to new inflation. Notice that $ r > \frac8{N} 
\simeq 0.16 $
for chaotic inflation while $  r < \frac8{N} $ for new inflation.}
\label{trirmulti}
\end{figure}

The upper and lower bounds on the inflaton mass ratio $ x = 10^6 \; 
\frac{m}{M_{Pl}} $
derived from eq.(\ref{dtrino}) take the same form as eq.(\ref{cotx}) 
\be \label{cotxh}
\frac{200 \, \sqrt3 \, \pi}{N} g_b(z_b) \; \sqrt{\Delta_-} < x 
< \frac{200 \, \sqrt3 \, \pi}{N} g_b(z_b) \; \sqrt{\Delta_+} \; ,
\ee
with the function,
\be \label{cotxh2}
g_b(z_b) = \frac{\kappa \; N \sqrt{z_b} \; |1 - 2  \, h  \, \sqrt{z_b} -z_b|
}{\left[w(h)- 2 \, z_b+ \frac83  \, h  \,  z_b^{3/2} + 
z_b^2\right]^{\frac32}}   \; .
\ee
Figs. \ref{trimasamulti}, 
\ref{trinsmulti} and \ref{trirmulti} depict respectively
the central value for $ x $ eq.(\ref{cotxh}), $n_s$  
eqs.(\ref{ntrino})-(\ref{nstrino})
and  $r$ eqs.(\ref{ntrino}) and (\ref{rtrino}) as functions
of $ y = \kappa \, N $ for different values of $ h < 0 $
for the trinomial potential eq. (\ref{trino}).
In all cases the spectrum turns to be red tilted ($n_s < 1$).

\begin{figure}[htbp]
{\epsfig{file=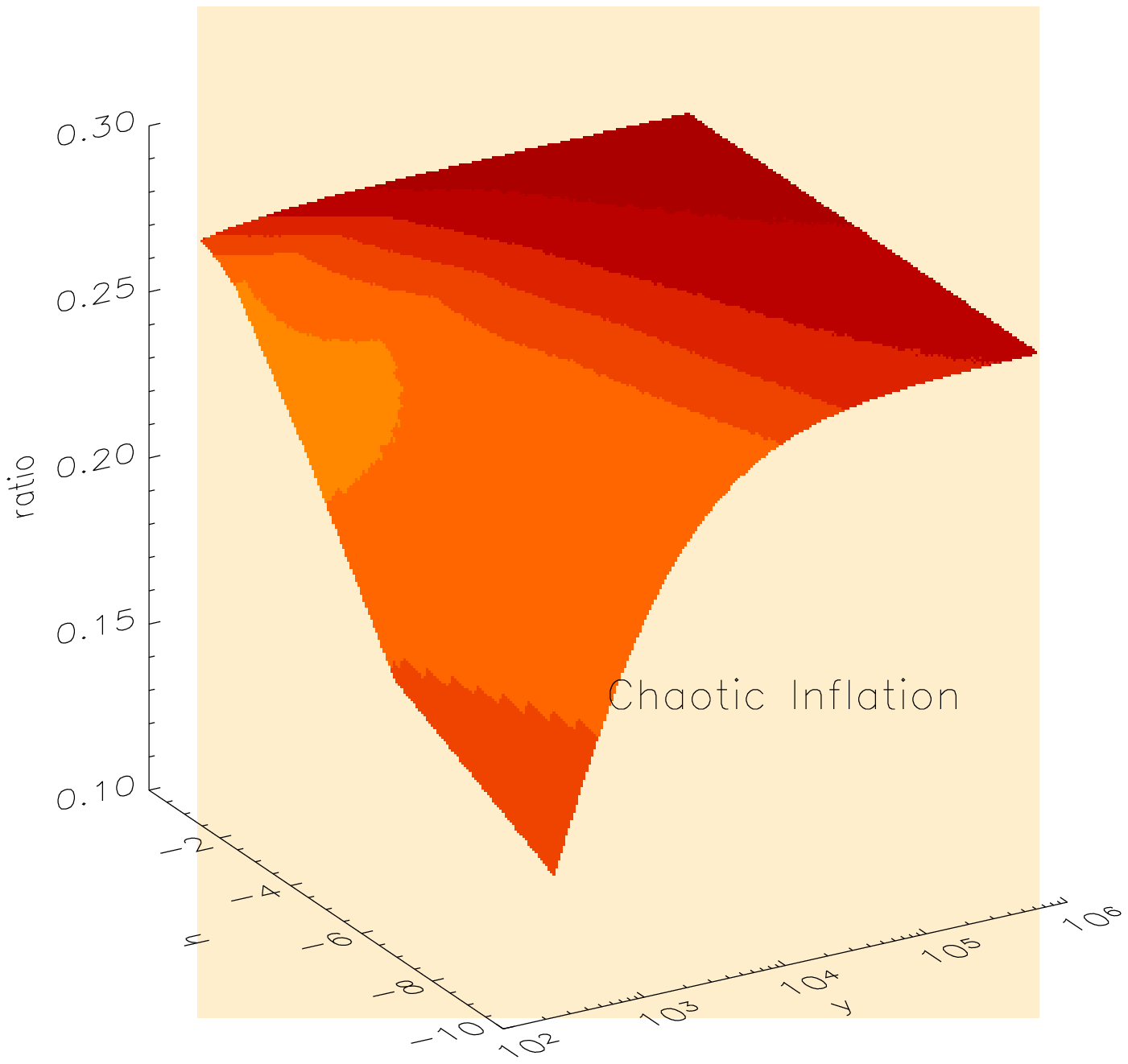,width=14cm,height=14cm}}
\caption{ $r$ as a function of $ y = \kappa \, N $  for $ N = 60$ and $ h < 0 $ 
with the trinomial potential eq. (\ref{trino})
as given by eqs.(\ref{ntrino}) and (\ref{rtrino}) for chaotic inflation.}
\label{trirsurfCAO}
\end{figure}

\begin{figure}[htbp]
{\epsfig{file=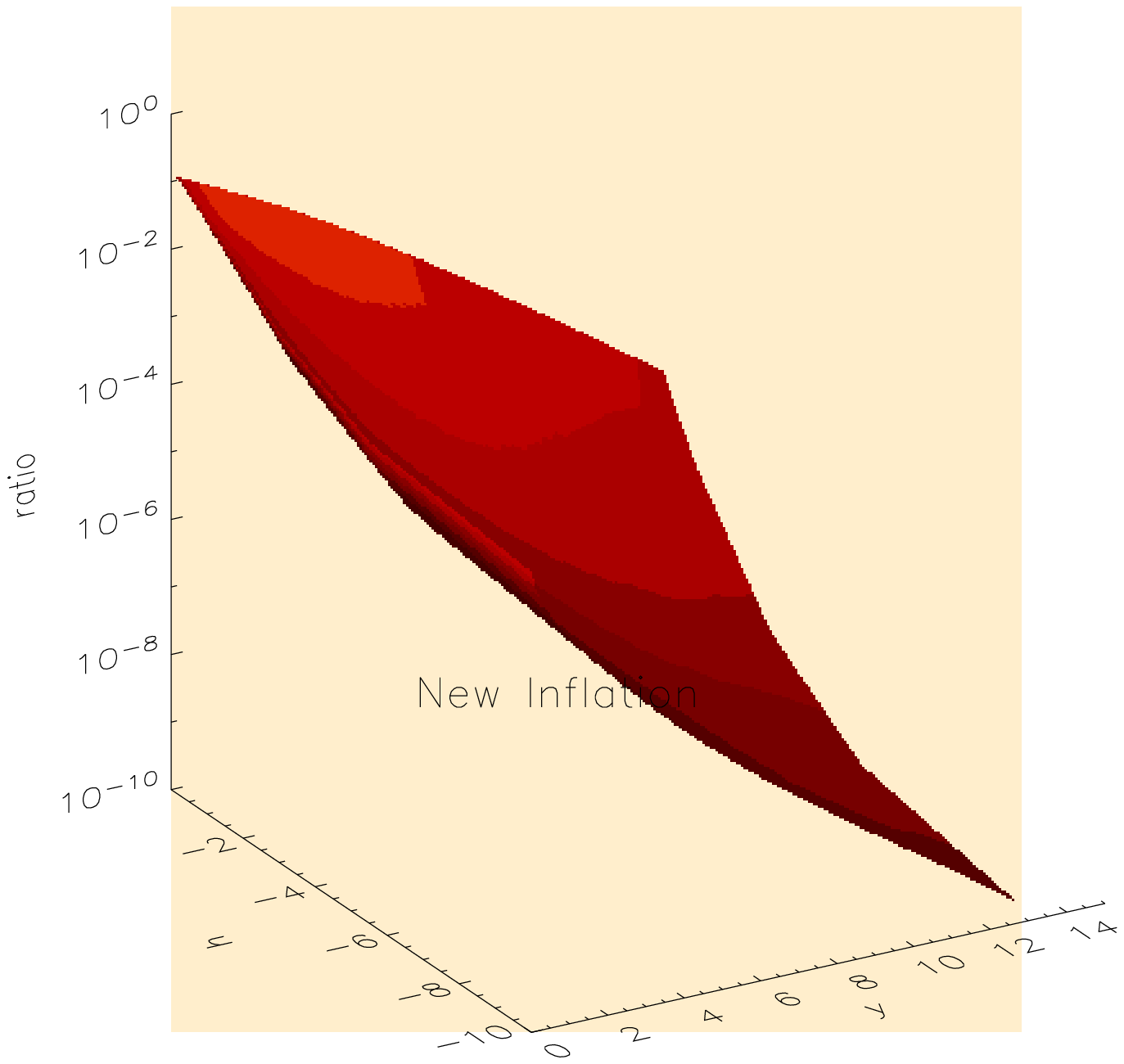,width=14cm,height=14cm}}
\caption{ $r$ as a function of $ y = \kappa \, N $  for $ N = 60$ and 
$ h < 0 $ with the trinomial potential eq. (\ref{trino})
as given by eqs.(\ref{ntrino}) and (\ref{rtrino}) for new inflation.}
\label{trirsurfNEW}
\end{figure}

The three-dimensional plots (\ref{trimasasurfCAO}, \ref{trinssurfCAO}, 
\ref{trirsurfCAO})
and (\ref{trimasasurfNEW}, \ref{trinssurfNEW}, \ref{trirsurfNEW}),
display $ x, \; n_s $ and $ r $ as functions of $ y = \kappa \, N $
and  $ h < 0 $, for chaotic and new inflation, respectively. 

\begin{figure}[htbp]
{\epsfig{file=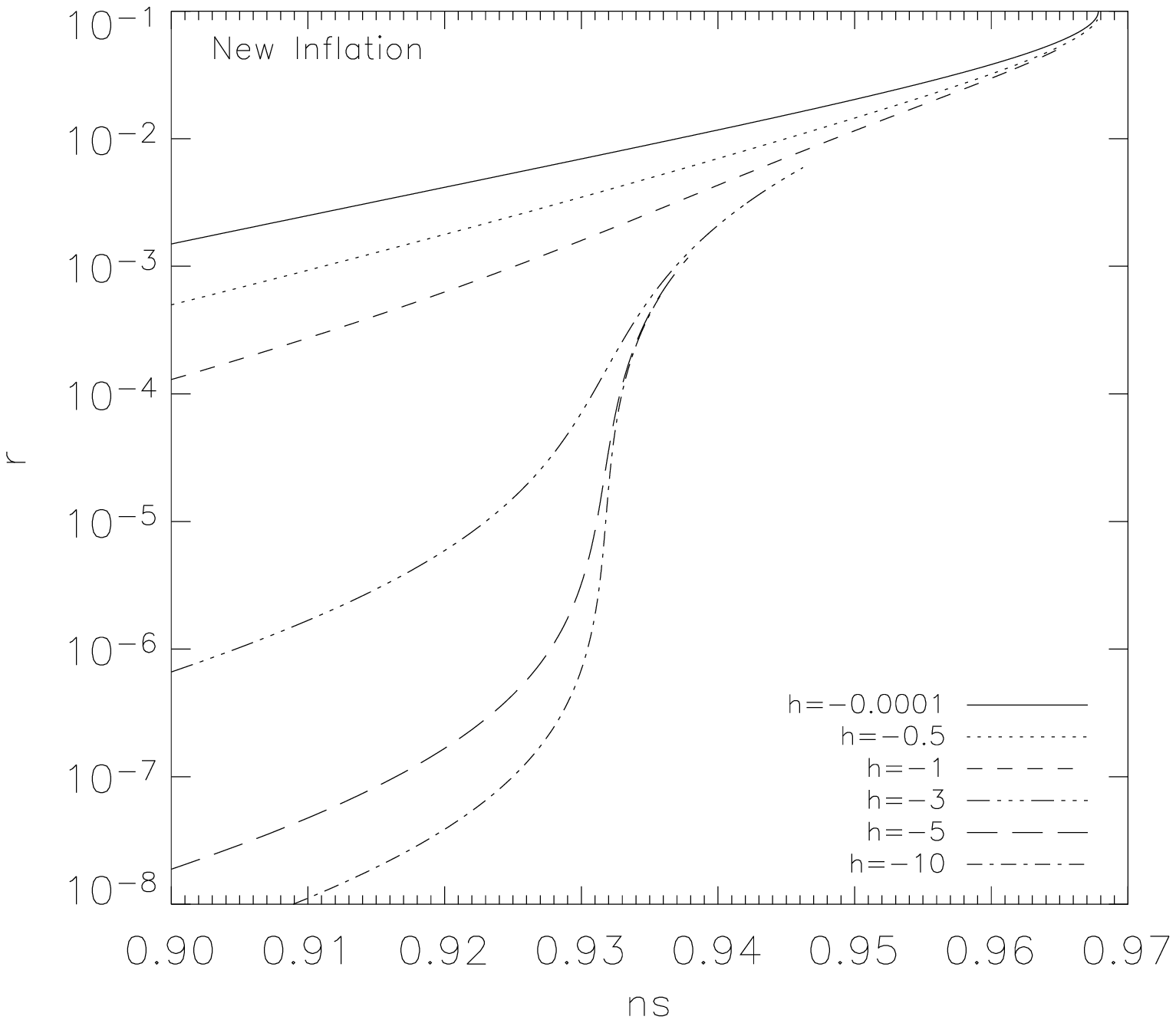,width=14cm,height=14cm}}
\caption{ $r$ as a function of $n_s$ for different 
values of $ h < 0 $ with the trinomial potential eq. (\ref{trino})
as given by eqs.(\ref{ntrino}), (\ref{nstrino}) 
and (\ref{rtrino}) for new inflation.}
\label{trinsrmulti}
\end{figure}

\begin{figure}[htbp]
{\epsfig{file=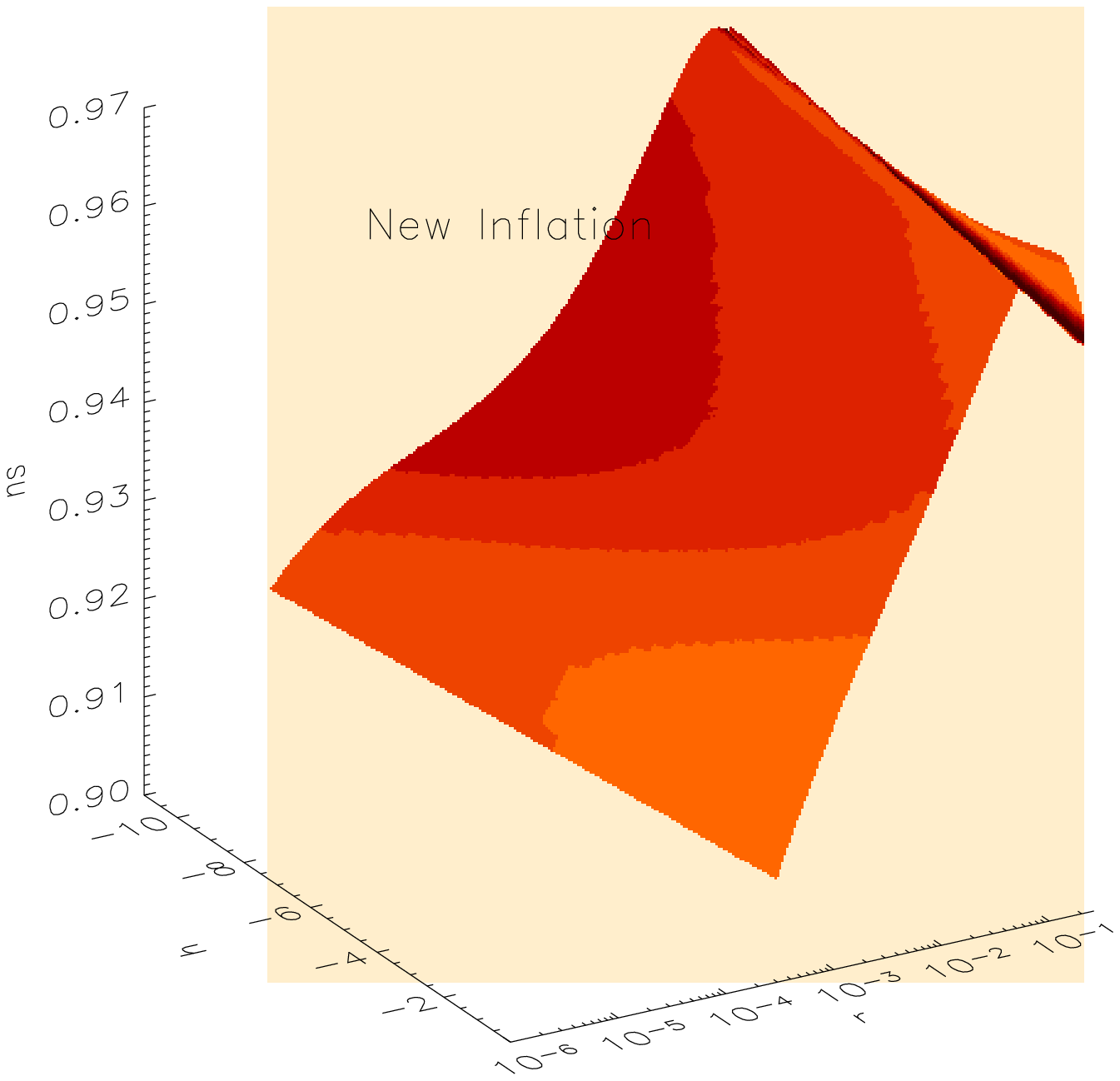,width=14cm,height=14cm}}
\caption{ $n_s$ as a function of $ r $ and $ h < 0 $ 
with the trinomial potential eq. (\ref{trino})
as given by eq.(\ref{nstrino})  and (\ref{rtrino}) for new inflation.}
\label{trinsrsurf}
\end{figure}

\begin{figure}[htbp]
{\epsfig{file=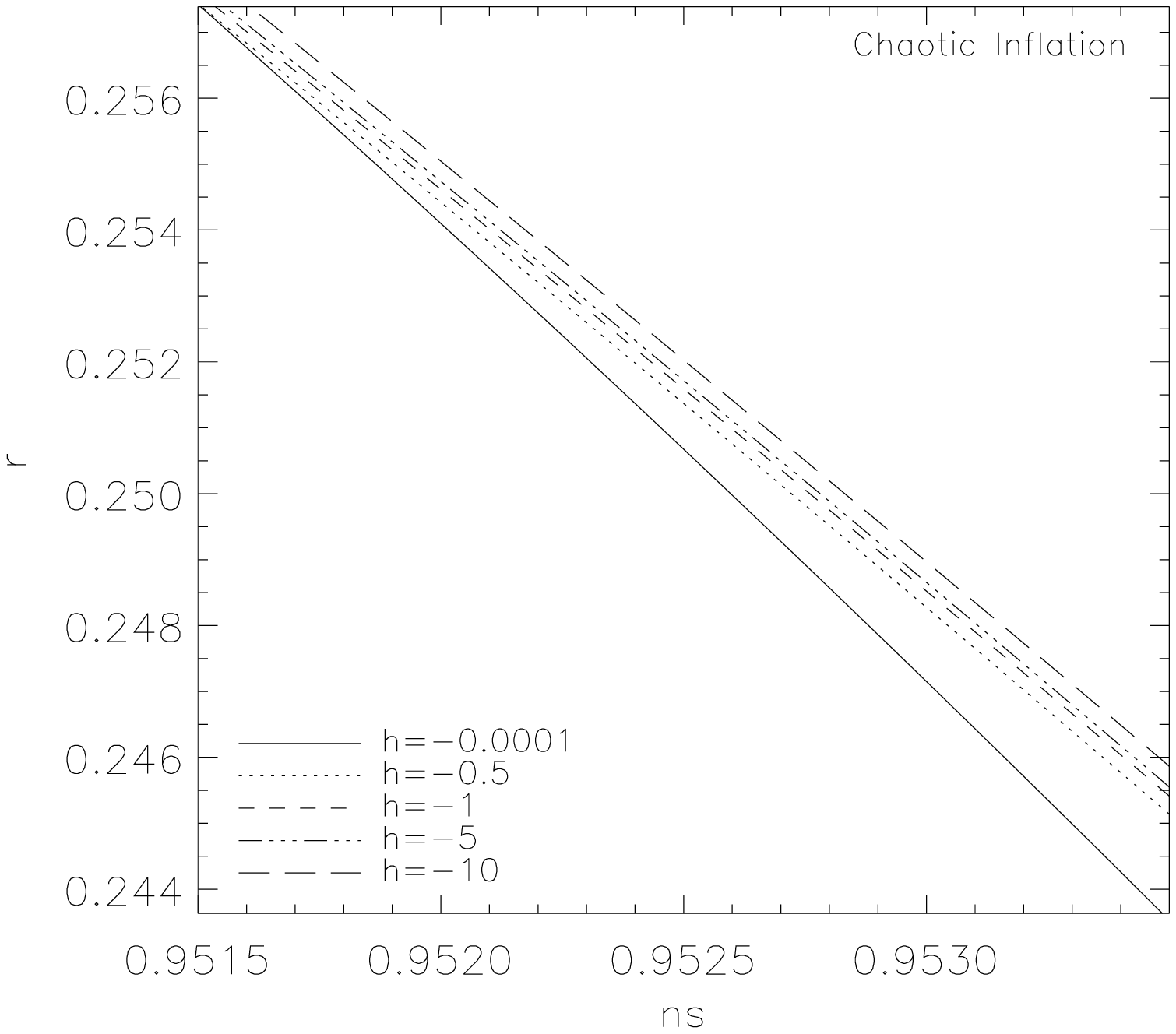,width=14cm,height=14cm}}
\caption{ $ r $ as a function of  $n_s$ for different 
values of $ h < 0 $ with the trinomial potential eq. (\ref{trino})
as given by eq.(\ref{nstrino})  and (\ref{rtrino}) for chaotic inflation.}
\label{trinsrcao}
\end{figure}

\begin{figure}[htbp]
{\epsfig{file=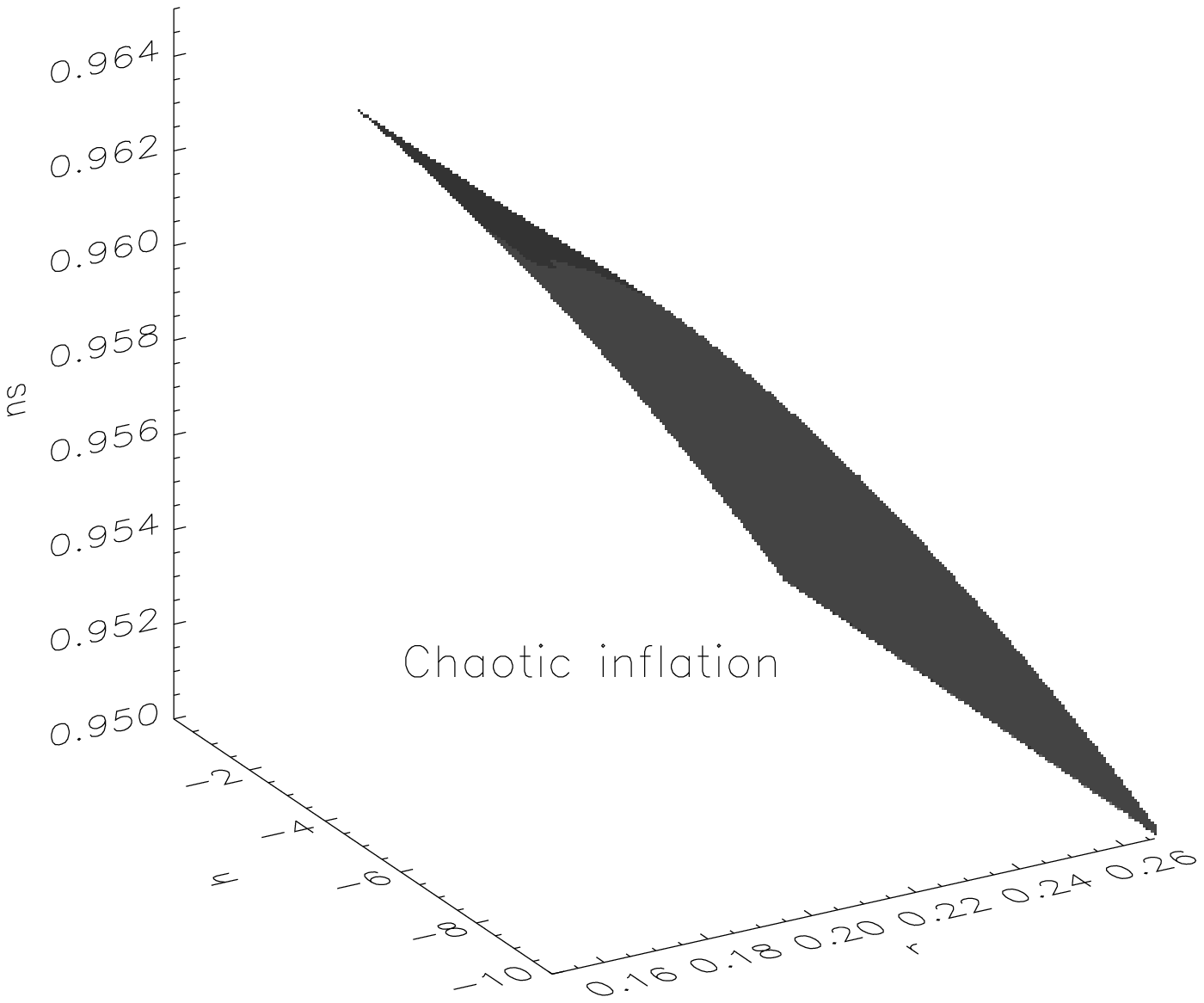,width=14cm,height=14cm}}
\caption{ $n_s$ as a function of $ r $ and $ h < 0 $ 
with the trinomial potential eq. (\ref{trino})
as given by eq.(\ref{nstrino})  and (\ref{rtrino}) for chaotic inflation.}
\label{trinsrsurfCAO}
\end{figure}

Fig. \ref{trinsrmulti} depicts $r$ as a function of $n_s$ for 
new inflation with the trinomial potential  eq. (\ref{trino})
for different values of $ h < 0 $ as given by eqs.(\ref{ntrino}) 
and (\ref{rtrino}). We see that $r$ {\bf decreases} when $ n_s $ goes below 
unit, as in the binomial case. In addition, $ r $ {\bf decreases} for 
increasing asymmetry $ |h| $ at a fixed $ n_s $ (with $ h<0 $). 
Therefore, the trinomial 
potential  eq. (\ref{trino}) can yield {\bf very small $r$} with  
{\bf $ n_s < 1 $ and near unit} for new inflation [see eq.(\ref{nsrtrikG})].
In this case we have the upper bound $ r \leq \frac{8}{N} \simeq 0.16 $ 
[see fig. \ref{trirmulti} and  eq.(\ref{cotrino})].
  
The three-dimensional plot fig. \ref{trinsrsurf} depicts $n_s$ as a function 
of $ r $ and $ h $ with the trinomial potential eq. (\ref{trino})
as given by eqs.(\ref{ntrino}), (\ref{nstrino}) 
and (\ref{rtrino}) for new inflation.

In figs. \ref{trinsrcao} and \ref{trinsrsurfCAO}
we plot $r$ as function of $n_s$ and $h$ for chaotic inflation for the 
trinomial potential eq. (\ref{trino}). We see that $r$ increases when $ n_s $ 
goes below unit, 
as in the binomial case. For chaotic inflation with the trinomial potential  
eq. (\ref{trino}) we have the lower bound $ r \geq \frac{8}{N} $ [see fig. 
\ref{trirmulti} and eq.(\ref{cotrino})]. Notice that fig. \ref{trinsrcao} 
displays a very small range
of variation for $ n_s $ and $ r $, actually,  $r$ as a function of $n_s$
has a very {\bf weak} dependence on the asymmetry $ h < 0 $ 
for chaotic inflation. 

\subsubsection{\bf Limiting Cases}

Let us now consider the limiting cases: the shallow limit  ($ \kappa \to 0 $)
and the steep limit $ \kappa \to \infty $.

In the shallow  limit 
$ \kappa \to 0 , \; \kappa \, N(z_b) $ tends to its minimum $ z_b = z_+ $.
We find from eqs.(\ref{nstrino})-(\ref{rtrino}),
\be\label{cotrino}
 n_s \buildrel{\kappa \to 0}\over= 1 - \frac{2}{N} \simeq 0.96  
\quad , \quad r \buildrel{\kappa \to 0}\over= \frac{8}{N}  \simeq 0.16 
\ee
which coincide with $ n_s $ and $ r $ eq. (\ref{monofi2}) for the monomial
quadratic potential. That is, the  $ \kappa \to 0 $ limit
is $h$-independent as expected since for fixed $h, \; \gamma 
\buildrel{\kappa \to 0}\over= {\cal O}(\sqrt{\kappa}) $ and the 
inflaton potential eq.(\ref{trino}) becomes purely quadratic.

\bigskip

In the  steep limit $ \kappa \to \infty $ we have two possibilities:
for new inflation $ z_b $ tends to zero while for chaotic inflation $ z_b $
tends to infinity. 

For new inflation we find from eq.(\ref{ntrino})
\be\label{trikG}
\kappa \, N(z_b)  \buildrel{z_b \to 0}\over=- w(h) \log z_b
-q(h) -1 + {\cal O}(\sqrt{z_b}) \quad \mbox{new~inflation} \; ,
\ee
where
$$
q(h) \equiv 2 \,  w(h) \log[\Delta-|h|] -
\frac23 \; \left( h^2 + |h| \; \Delta \right) 
\left\{ 8 \, \Delta^2 \, 
\log\left[\frac12 \left(1 - \frac{|h|}{\Delta}\right)\right] - 1 \right\}
\; ,
$$
is a monotonically increasing function of the asymmetry 
$ |h| : \; 0 \leq q(h) < \infty $ for $ 0 < |h| < \infty $.

Then, eqs.(\ref{nstrino})-(\ref{rtrino}) yield,
\be\label{nsrtrikG}
n_s \buildrel{\kappa \gg 1}\over=1 - \frac{\kappa}{w(h)}
\quad , \quad r \buildrel{\kappa \gg 1}\over=
16 \;   \frac{\kappa}{w^2(h)} \, e^{- \frac{\kappa 
\, N + 1+q(h)}{w(h)} }\quad 
\mbox{new~inflation}
\ee
In the $ h \to 0 $ limit we recover from eqs.(\ref{trikG})-(\ref{nsrtrikG})
the results for new inflation with a purely quartic potential 
eq.(\ref{nsrpuro4}) since $ w(0) = 1$ and $ q(0) = 0 $.

We find for the function $ g_b(z_b) $ governing the inflaton mass ratio
eq. (\ref{cotxh2})
\be\label{gzasi}
g_b(z_b) \buildrel{\kappa \gg 1}\over=
\frac{\kappa \; N}{w^{\frac32}(h)} \; e^{- \frac{\kappa 
\, N + 1+q(h)}{2 \, w(h)} } \; , 
\quad \mbox{new~inflation}
\ee
or using eq.(\ref{nsrtrikG}),
\be\label{gzasi2}
g_b(z_b) \buildrel{\kappa \gg 1}\over= \frac{N}{4} \; \sqrt{r(1-n_s)} \; .
\ee
This implies for the mass ratio bounds in eq.(\ref{cotxh})
$$
50 \, \sqrt3 \, \pi \; \sqrt{r(1-n_s)} \; \sqrt{\Delta_-} < x 
< 50 \, \sqrt3 \, \pi \; \sqrt{r(1-n_s)} \; \sqrt{\Delta_+} \; .
$$
or
\be\label{cotlim}
x = 127 \; \sqrt{r(1-n_s)} \pm 6 \% \; .
\ee
We find in this limiting case that the mass ratio is {\bf directly} related
to the observable quantities $ n_s $ and $ r $. For example, if 
$ 1-n_s \sim r \sim 10^{-n} $, then $ x \sim 10^{2-n} $.

Notice that in all cases the values of $n_s - 1 $ and $ r $ are of the order
$ \frac1{N} $. 

\bigskip 

We read from figs. \ref{trinsmulti} and \ref{trirmulti} the behaviour of 
$ n_s $ and $ r $ in chaotic and new inflation ($ h < 0 $):
\begin{itemize}
\item{ Both for chaotic and new inflation $n_s$ decreases with the steepness
$ \kappa $ for fixed asymmetry $ h < 0 $ and grows with the asymmetry 
$ |h| $ for fixed steepness $ \kappa $.}

\item{For chaotic inflation $ r $  grows with the steepness $ \kappa $ 
for fixed asymmetry  $ h < 0 $
and decreases with the asymmetry  $ |h| $ for fixed steepness $ \kappa $.
For new inflation  $ r $ does the opposite: it decreases with the steepness 
$ \kappa $ for fixed asymmetry  $ h < 0 $ while it grows with the asymmetry  
$ |h| $ for fixed steepness $ \kappa $.}

\end{itemize}

\bigskip

From figs. \ref{trimasamulti}, \ref{trinsmulti} and \ref{trirmulti}
we can understand how the mass ratio $ \frac{m}{M_{Pl}} $ varies with $ n_s $ 
and $ r $. For chaotic inflation we have $ r \geq \frac{8}{N} $ and $ 10^5 \; 
\frac{m}{M_{Pl}} $ stays larger than unit in order to keep $ n_s $ in the 
observed WMAP range. 
For new inflation, we have  $ r \leq \frac{8}{N} $ and  $ \frac{m}{M_{Pl}} $
decreases for decreasing $ r $. However, if we consider 
$ 10^5 \; \frac{m}{M_{Pl}} < 1 $ say, eq.(\ref{cotlim}) and  
figs. \ref{trimasamulti}, \ref{trinsmulti} and \ref{trirmulti}
show that $ (1-n_s) \, r $ must be $ \sim 10^{-6} $ which is probably a
too small value.
That is, there is a limiting value for the inflaton mass $ x_0 \equiv 10^6  \; 
\frac{m_0}{M_{Pl}} \simeq 0.1 $ such that $ m_0 \simeq  10^{-7} \; M_{Pl} $ is
a {\bf minimal} inflaton mass in order to keep $ n_s $ and $ r $ within the 
WMAP data.

\bigskip

This concludes our discussion of the trinomial inflaton potential
with $ m^2 < 0 $. The case  $ m^2 > 0 $ can be treated analogously
and yields results conceptually similar to the binomial potential:
one finds chaotic inflation modulated by the asymmetry parameter $ h $.
We shall not consider such case here since it yields, as always
in chaotic inflation, a ratio $ r > \frac8{N} $. 

\section{Hybrid Inflation}

In the inflationary models of hybrid type, the inflaton is coupled to another 
scalar field $ \sigma_0 $ through a potential of the type\cite{lin}
\bea\label{Vhib1}
&&V_{hyb}(\phi,\sigma_0) = \frac{m^2}{2} \; \phi^2 + \frac{g_0^2}{2} \; 
\phi^2 \; \sigma_0^2 + \frac{\mu_0^4}{16 \, \Lambda_0} 
\left(\sigma_0^2 -  \frac{4 \, \Lambda_0}{\mu_0^2} \right)^2= \cr \cr
&&  =\frac{m^2}{2} \; \phi^2 + \Lambda_0+\frac12 \; (g_0^2 \; 
\phi^2-\mu_0^2) \;  \sigma_0^2+\frac{\mu_0^4}{16 \; \Lambda_0} \; 
\sigma_0^4\; .
\eea
where $ \mu_0^2 > 0 $ is of the order $ m^2 > 0 , \; \Lambda_0 > 0 $ plays 
the role of a cosmological constant and $ g_0^2 $  couples $ \sigma_0 $ with 
$ \phi $.

\medskip

The initial conditions are chosen such that $\sigma_0$ and $ \dot\sigma_0$ 
are very small and one can therefore consider,
\be\label{vsig0}
V_{hyb}(\phi,0) =\frac{m^2}{2} \; \phi^2 + \Lambda_0 \; .
\ee
One has then inflation driven by the cosmological constant $ \Lambda_0 $ in 
the regime $  \phi(0) \ll \sqrt\Lambda_0$. The inflaton field $ \phi(t) $ 
decreases with time while the scale factor $ a(t) $ grows exponentially.
We see from eq.(\ref{Vhib1}) that
$$
m_{\sigma}^2 = g_0^2 \; \phi^2 - \mu_0^2
$$
plays the role of a effective classical mass square for the field $ \sigma_0 $.
The initial value of $ m_{\sigma}^2 $ depends on the initial conditions
but is typically positive.
Anyway, since in chaotic inflation the inflaton field $ \phi $ decreases
with time, $ m_{\sigma}^2 $ will be necessarily negative at some 
moment during inflation.
At such moment, spinodal (tachyonic) unstabilities appear and the
field $ \sigma $ starts to grow exponentially till it dominates the
energy of the universe. Inflation stops at such time and then after, 
a matter dominated regime follows.

Hybrid inflation can also be obtained with other couplings like
\be\label{Vhib2}
{\tilde V}_{hyb}(\phi,\sigma_0) =  \frac{m^2}{2} \; \phi^2 + 
\Lambda_0+\frac12 \; (2 \; m \; g_0 \; 
\phi-\mu_0^2) \;  \sigma_0^2+\frac{\tilde g}4 \; \sigma_0^4\; .
\ee
where $ g_0 $ couples $ \sigma_0 $ with $ \phi $,  and the quartic coupling
$ {\tilde g} > 2 \,  g_0^2 $ ensures the stability of the model.
Here, the effective classical mass square for $ \sigma_0 $ reads,
$$
m_{\sigma}^2 = 2 \; m \; g_0 \; \phi - \mu_0^2 \; .
$$
Again $ m_{\sigma}^2 $ becomes negative at some moment of inflation due
to the fact that the inflaton field $ \phi $ decreases with time.
Spinodal unstabilities are then triggered and inflation eventually
stops followed by a matter dominated regime.
As for the potential $ V_{hyb}(\phi,\sigma_0) $, in the 
regime $  \phi(0) \ll \sqrt\Lambda_0$,  inflation is driven by the 
cosmological constant $ \Lambda_0 $. 

\bigskip

In hybrid inflation the role of the field $ \sigma_0 $ is to stop inflation.
This field is negligible when the relevant cosmological scales cross out the 
horizon. Hence, $ \sigma_0 $ does not affect the spectrum of density and
tensor fluctuations except through the number of efolds. 

\bigskip

The evolution equations in dimensionless variables for the model
 $ V_{hyb}(\phi,\sigma_0) $  eq.(\ref{Vhib1}) take the form
\bea\label{ecmovh}
&&h^2(\tau) = \frac{2}{3} \; \left[ {\dot \varphi}^2 + \varphi^2 
+ \frac{\mu^4}{4 \; \Lambda}\left(\sigma^2 - \frac{2 \, \Lambda}{\mu^2} 
\right)^2 + g^2 \; \sigma^2 \; \varphi^2
\right] \; , \cr \cr
&&{\ddot \varphi} + 3 \, h \, {\dot \varphi} + \varphi + g^2 \; \sigma^2 \; 
\varphi = 0 \; , \label{fihib} \\
&&{\ddot \sigma} + 3 \, h \, {\dot \sigma} - \mu^2 \; \sigma + g^2 \; \sigma 
\; \varphi^2 +\frac{r^2}{2 \; \Lambda} \; \sigma^3 =0 \; . \nonumber
\eea
where 
$$ \sigma(\tau) \equiv \frac{\sigma_0(t)}{M_{Pl}} \quad , \quad
 g^2 \equiv g_0^2 \; \frac{M_{Pl}^2}{m^2}\quad , \quad
\mu^2 \equiv \frac{\mu_0^2}{m^2} \quad  \mbox{and}  
\quad \Lambda \equiv \frac{2 \, \Lambda_0}{m^2 \;M_{Pl}^2} \; .
$$
In the slow-roll and $ \Lambda$-dominated  regime:
$$
{\dot \phi }(0) \ll m \, \phi(0) \ll  \sqrt\Lambda \; ,
$$
we can neglect the field $\sigma$ and approximate the evolution equations 
(\ref{fihib}) by
\bea
&& 3 \, h \, {\dot \varphi} + \varphi = 0\; , \cr \cr
&&h^2(\tau) = \frac{2}{3} \; \left[ \varphi^2 + \Lambda  \right] \; .
\eea
The number of efolds from the time $\tau$ till the end of inflation
is then given by eq.(\ref{nef}),
\be\label{Nhib}
N(\tau) = \int_0^{\tau} h(\tau) \; d\tau =  -  \int_{\varphi(\tau)}^{\varphi_0}
\frac{v(\varphi)}{v'(\varphi)} \; d\varphi
= \frac14 \left[ \varphi^2(\tau)- \varphi^2_0 \right]
+ 2 \, \Lambda \; \log\frac{\varphi(\tau)}{\varphi_0}
\ee
where $ \varphi_0 $ is the inflaton field by the end of inflation.
We have verified this approximation by integrating numerically 
eqs. (\ref{fihib}). 

We see that the field and its dynamics only appears in eq.(\ref{Nhib})
through the value of $ \varphi_0 $ where inflation stops. The
value of $ \varphi_0 $ follows by solving eqs.(\ref{ecmovh}) and 
depends on the initial conditions as well as on the parameters 
$ g, \; \mu $ and $ \Lambda $.

For $ \Lambda \to 0 $ hybrid inflation becomes chaotic inflation with
the monomial potential $ \frac12  \varphi^2$. In that limit eq.(\ref{Nhib})
becomes eq.(\ref{znr}) with $ \kappa \to 0 $ as it should be. 

\medskip

The spectral indices are given by eqs.(\ref{indesp})-(\ref{epseta}) and
the amplitude of adiabatic perturbations by eq.(\ref{amplis}).
By using the potential eq.(\ref{vsig0}) in dimensionless variables we find,
\bea\label{indhi}
&&v(\varphi,0) = \frac12 \, ( \varphi^2 + \Lambda) \quad , \quad
\eta =\frac{2}{\varphi^2 + \Lambda} \quad ,  \quad \epsilon= 
\frac{2 \, \varphi^2}{(\varphi^2 + \Lambda)^2}\; , \\ \cr
&&|{\delta}_{k\;ad}^{(S)}|^2 = \frac{1}{96 \, \pi^2 } \left( \frac{m}{M_{Pl}} 
\right)^2 \frac{(\varphi^2 + \Lambda)^3}{\varphi^2} \quad ,  \quad 
n_s = 1 + 4 \; \frac{\Lambda - 2 \; \varphi^2}{(\varphi^2 + \Lambda)^2}
\quad ,  \quad r = 32 \; \frac{\varphi^2}{(\varphi^2 + \Lambda)^2} 
\label{nshib} \; .
\eea
where $ \varphi $ is the inflaton at the moment of the first horizon crossing.

\medskip

\begin{figure}[htbp]
{\epsfig{file=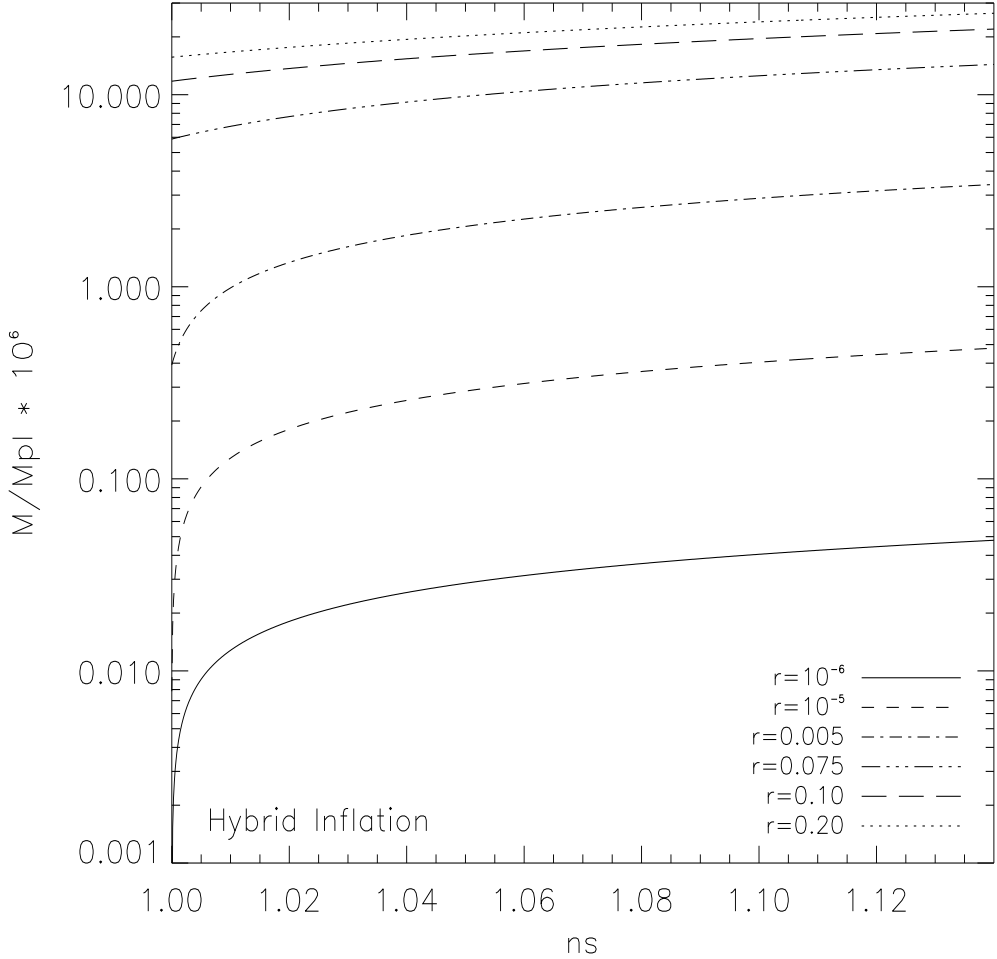,width=14cm,height=14cm}}
\caption{ $ x = 10^6 \; \frac{m}{M_{Pl}} $ as a function of $ n_s $ for 
different values of $r$ in hybrid inflation according to eq. (\ref{xhi}).} 
\label{hmasans}
\end{figure}

\begin{figure}[htbp]
{\epsfig{file=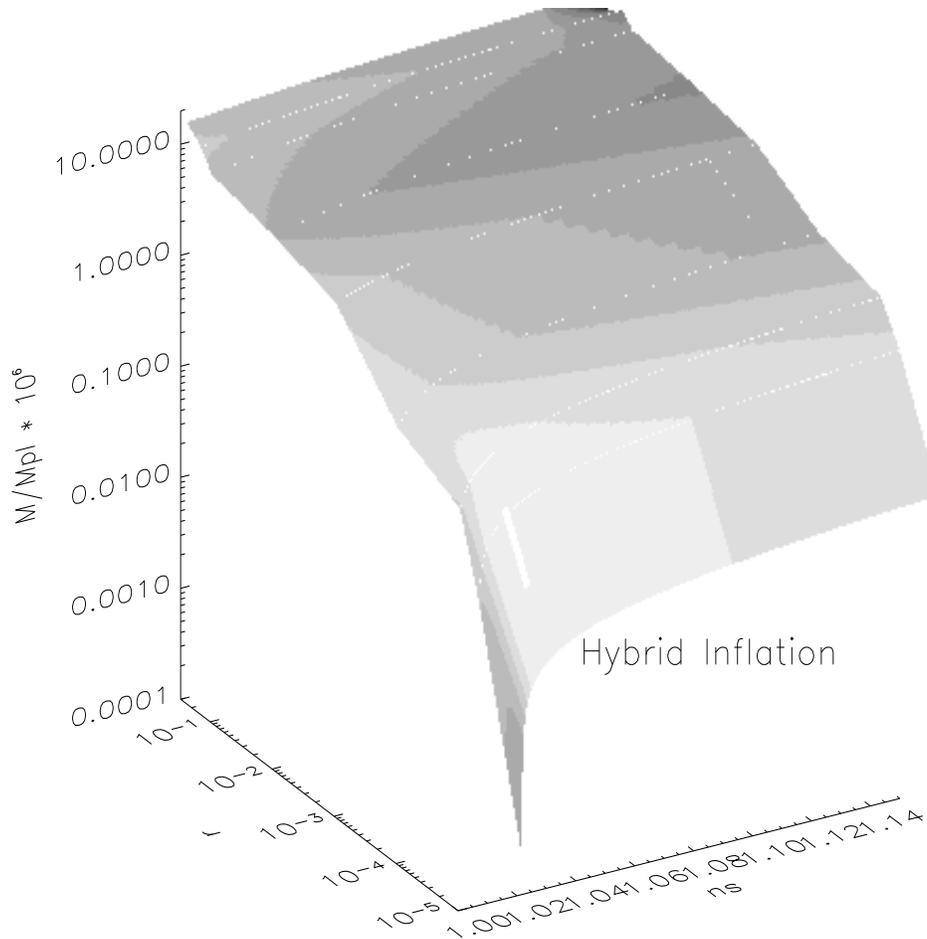,width=14cm,height=14cm}}
\caption{ $ x = 10^6 \; \frac{m}{M_{Pl}} $ as a function of $ n_s $ and $r$
in hybrid inflation according to eq. (\ref{xhi}) } \label{hymass}
\end{figure}

\begin{figure}[htbp]
{\epsfig{file=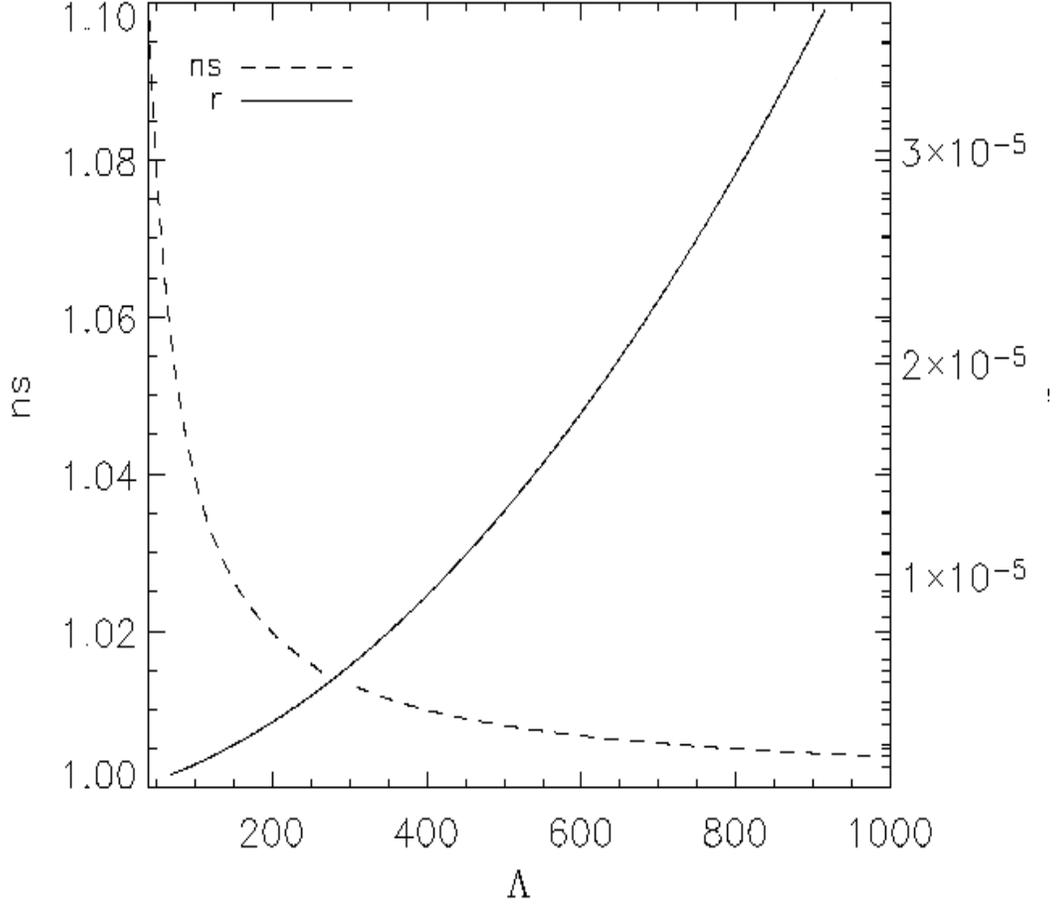,width=14cm,height=14cm}}
\caption{  $ n_s $ and $r$ as functions of $ \Lambda $ for a fixed $x$
according to eqs. (\ref{r}) and  (\ref{ns}). Here, $ x = 0.0411 $.} 
\label{hyrns}
\end{figure}

Since $ \varphi \ll \sqrt \Lambda $ eq.(\ref{indhi})-(\ref{nshib}) simplify as,
\bea\label{indhis}
&&\eta =\frac{2}{\Lambda} \quad ,  \quad \epsilon= 
2 \, \left(\frac{\varphi}{\Lambda}\right)^2
\quad ,  \quad 
|{\delta}_{k\;ad}^{(S)}|^2 = \frac{1}{96 \, \pi^2 } \left( \frac{m}{M_{Pl}} 
\right)^2 \left(\frac{\Lambda}{\varphi}\right)^2 \; \Lambda \; , \cr \cr
&& 
n_s = 1 + \frac4{\Lambda} - 12\; \frac{\varphi^2}{\Lambda^2} 
\quad ,  \quad r = 32\; \frac{\varphi^2}{\Lambda^2}  \; .
\eea
In terms of the variable $ x = 10^6 \; \frac{m}{M_{Pl}} $ [eq.(\ref{x})] 
we get
\be\label{r}
\frac{\varphi^2}{\Lambda^2} = \frac{1}{96 \, \pi^2}
\frac{x^2}{|{\delta}_{k\;ad}^{(S)}|^2} 10^{-12} \; \Lambda
= 0.478 \times 10^{-6} \; x^2 \; \Lambda
\quad ,  \quad 
\frac{r}{\Lambda} = \frac{1}{3 \, \pi^2}\frac{x^2}{|{\delta}_{k\;ad}^{(S)}|^2}
 10^{-12} = 1.52 \; 10^{-5} \; x^2 \quad ,  
\ee
where we used the WMAP data for the amplitude of scalar perturbations 
eq. (\ref{ampmap}).
Since,
\be\label{ns}
n_s = 1 + \frac{4}{\Lambda} - \frac38 \; r \; ,
\ee
we find from eqs.(\ref{r}) and (\ref{ns}),
\bea\label{xhi}
&&x =  10^6 \; \frac{m}{M_{Pl}} =  5 \, \pi \, \sqrt{3} \; 10^5 \, 
|{\delta}_{k\;ad}^{(S)}| \; \sqrt{r(1-n_s)} = 127 \, 
\sqrt{r \, \left(n_s -1 + \frac38 \; r\right) } \; , \cr \cr
&& \frac{\Lambda_0^{\frac14}}{M_{Pl}} = 0.0135 \; r^{\frac14} 
\qquad , \qquad \frac{\Lambda_0}{M^4_{Pl}} = 0.329 \times 10^{-7} \; r \; .
\eea
Notice that the expression for the mass ratio has a similar structure than
for new inflation (in the limiting case) eq.(\ref{cotlim}).
In figs. \ref{hmasans}-\ref{hymass} we plot $ x = 10^6 \; \frac{m}{M_{Pl}} $ 
as a function of $ n_s $ and $r$ according to eq. (\ref{xhi}).
We see that $ \frac{m}{M_{Pl}} $ {\bf decreases} when $ r $  {\bf and} 
$ n_s - 1 $ both approach zero. 
Fig. \ref{hyrns} displays  $ n_s $ and $r$ as functions of $ \Lambda $ for a 
fixed $x$ according to eqs. (\ref{r}) and  (\ref{ns}).

\bigskip

From eq.(\ref{nshib}) we obtain $ \varphi^2 $ as a function of $n_s$ and 
$\Lambda$,
\be
\varphi^2_{\pm} = - \Lambda +\frac4{n_s-1}\left[ -1 \pm \sqrt{1 + \frac34 \; 
\Lambda \; (n_s-1)} \right] \; .
\ee
We see that $ \varphi^2_+ $ is positive and hence physical for $ n_s>1 $ and 
$ \Lambda < \frac4{n_s-1} $
while  $ \varphi^2_- $  is positive and hence physical for $ n_s<1 $ and 
$ \Lambda < \frac4{3|n_s-1|} $. Hence,  the value of $ n_s-1 $ gives an  {\bf 
upper bound} on the cosmological constant $\Lambda$. 
In addition, one sees from  eq.(\ref{nshib}) that we have for hybrid inflation
$$ 
n_s-1 - \frac4{\Lambda} < 0 \quad \mbox{for~all~} \varphi^2 > 0 \; .
$$
Moreover, when $ \Lambda \gg \varphi^2 $ ($\Lambda$-dominated regime),
we see from eq. (\ref{nshib}) that the spectrum exhibits a blue tilt 
($ n_s > 1 $).
[Both chaotic and new inflation yield red tilted spectra ($  n_s < 1 $) 
as discussed in sec. III].

We see from eqs.(\ref{Nhib}) and (\ref{indhis}) that 
$ \Lambda \sim N $ and hence
$$ n_s - 1 = {\cal O}\left(\frac1{N}\right) \quad , \quad
r  = {\cal O}\left(\frac1{N^2}\right) \; .
$$

\section{Conclusions and Implications for Supersymmetry and String Theory}

Setting the inflaton mass $m=0$ in polynomial potentials like eq.(\ref{V}) 
implies a highly particular non-generic choice. 
WMAP \cite{WMAP} unfavours such a choice and supports a generic polynomial 
potential.  Actually, we find that the fact that the pure $ \phi^4 $ 
potential is disfavoured implies a {\bf lower bound} on $m$. As discussed 
in secs. III and IV one gets $ m \gtrsim 10^{13}$GeV.

The potential which {\bf best fits} the present data for red tilted spectrum 
($ n_s < 1 $) and which {\bf best prepares} the way to the expected data 
(a small 
$ r \lesssim 0.1 $) is given by the trinomial potential eq.(\ref{potint}) with
a negative  $ \varphi^2 $ term, that is {\bf new} inflation. 
In new inflation we have the upper bound $ r \leq \frac{8}{N} \simeq  0.16 $.

\bigskip

The data on the spectral indices should be able to make soon a clear selection
between inflationary models: we see that a measured upper bound 
$ r \lesssim 0.16 $ 
excludes chaotic inflation. If this happens to be the case, then 
whether $n_s$ turns to be above or below unit will
exclude either new or hybrid inflation, respectively. 

\bigskip

The grand unification idea consists in that at some energy scale all three
couplings (electromagnetic, weak and strong) should become of the same
strength.  In this case, such grand unified scale turns out to be  $ E
\sim 10^{16}$GeV \cite{gut,sw}. 
The running of the couplings with the energy (or the length) is governed
 by the renormalization group.
For the standard model of electromagnetic, weak and strong interactions,
the renormalization group yields 
that the three couplings get unified approximately at $ \sim 10^{16}$GeV. 
A better convergence is obtained in supersymmetric extensions of the
standard model \cite{gut,sw}.

Neutrino oscillations and neutrino masses are currently explained in the 
see-saw mechanism as follows\cite{ita},
$$
\Delta m_{\nu} \sim \frac{M^2_{Fermi}}{M} 
$$
where $ M_{Fermi} \sim 250$ GeV is the Fermi mass scale, 
$M \gg  M_{Fermi} $ is a large energy scale and $ \Delta m_{\nu} $ 
is the difference between the neutrino masses for different flavors. 
The observed values for $ \Delta m_{\nu} \sim 0.009 - 0.05 $ eV naturally 
call for a mass scale $ M \sim 10^{15-16}$ GeV close to the GUT 
scale\cite{ita}. 

\bigskip

Eq.(\ref{vgen}) for the inflaton potential resembles the
moduli potential coming from supersymmetry breaking,
\be\label{susy}
V_{susy}(\phi) =  m_{susy}^4 \; v\!\left(\frac{\phi}{M_{Pl}}\right) \; ,
\ee
where $ m_{susy} $ stands for the supersymmetry breaking scale. 
Potentials with such form were used in the inflationary 
context in refs.\cite{susy}. In our context, eq.(\ref{susy}) implies that 
$  m_{susy} \sim 10^{16}$ Gev. That is, the susy breaking scale
$ m_{susy} $ turns out to be at the GUT scale $ m_{susy} \sim M_{GUT} $.

We see that the mass scale of the inflaton $ m \sim 10^{13}$GeV
can be related with $ M_{GUT} $ by a see-saw style relation,
\be\label{siso}
m  \sim \frac{M^2_{GUT}}{M_{Pl}} \; .
\ee
As discussed in secs. I and II the inflaton describes a condensate in a GUT
theory in which it may describe fermion-antifermion pairs.
Current identifications in the 
literature of such condensate field with a {\bf given}
fundamental field in a SUSY or SUGRA model  have so far no solid basis.
Moreover, the number of supersymmetric models is so large
that there is practically no way to predict which is {\bf the} correct model 
\cite{ramo}.

\bigskip

In order to generate inflation in string theory one needs 
first to generate a mass scale like the inflaton mass $m \sim 10^{13}$GeV. 
Such scale {\bf is not} present in the string action, neither in the action 
of the effective background fields (dilaton, graviton, antisymmetric tensor) 
which are massless. Without the presence
of the mass scales $m$ and $ M_{GUT} $ [related through
eq.(\ref{siso})],  there is {\bf no} hope in string theory 
to get a correct inflationary cosmology
describing the {\bf observed} CMB fluctuations\cite{noscu}.
Such scale should be generated 
dynamically perhaps from the string vacuum(ua) but this is still
an open problem far from being solved\cite{noscu}.
Actually, the very same problem hinders the derivation
of a GUT theory and the generation of the GUT scale from string theory.

\bigskip

Since no microscopic derivation of an inflationary model from a GUT is
available so far, it would seem too ambitious at this stage 
to look for a microscopic 
derivation of inflation from string theory. The derivation of
an inflationary cosmology reproducing the observed 
CMB fluctuations is at present too far away in string theory. 
However, an {\bf effective}
description of inflation in  string theory (string matter plus massless
backgrounds) could be at reach\cite{noscu}.

\end{document}